\documentclass{emulateapj} \usepackage{apjfonts, natbib} 
\usepackage{graphics, multirow} \usepackage{amsbsy}

\def\bicep{{\sc Bicep}}
\def\bolocam{{\sc Bolocam}}

\def\acbar{{\sc Acbar}}
\def\QUAD{{\sc QUaD}}

\def\boom{{\sc Boomerang}}

\def\spt{{\sc SPT}}

\def\camb{{\tt CAMB}}

\def\synfast{{\tt synfast}}

\def\spice{{\tt Spice}}

\def\deg{^\circ}

\def\lcdm{$\Lambda$CDM}

\def\aap{Astr. \& Astroph.\ }

\def\apjl{Astrophys.\ J.\ Lett.\ }
\def\apjs{Astrophys.\ J.\ Suppl.\ }

\shorttitle{\bicep\ instrument characterization}
\shortauthors{Takahashi et al.}
\submitted{Astrophysical Journal 711 (2010) 1141--1156}

\begin{document}

\title{Characterization of the BICEP Telescope for High-Precision \\
Cosmic Microwave Background Polarimetry}

\author{Y.~D.~Takahashi\altaffilmark{1}}
\author{P.~A.~R.~Ade\altaffilmark{2}}
\author{D.~Barkats\altaffilmark{3,4}}
\author{J.~O.~Battle\altaffilmark{5}}
\author{E.~M.~Bierman\altaffilmark{6}}
\author{J.~J.~Bock\altaffilmark{3,5}}
\author{H.~C.~Chiang\altaffilmark{3,7}}
\author{C.~D.~Dowell\altaffilmark{5}}
\author{L.~Duband\altaffilmark{8}}
\author{E.~F.~Hivon\altaffilmark{9}}
\author{W.~L.~Holzapfel\altaffilmark{1}}
\author{V.~V.~Hristov\altaffilmark{3}}
\author{W.~C.~Jones\altaffilmark{3,7}}
\author{B.~G.~Keating\altaffilmark{6}}
\author{J.~M.~Kovac\altaffilmark{3}}
\author{C.~L.~Kuo\altaffilmark{10}}
\author{A.~E.~Lange\altaffilmark{3}}
\author{E.~M.~Leitch\altaffilmark{11}}
\author{P.~V.~Mason\altaffilmark{3}}
\author{T.~Matsumura\altaffilmark{3}}
\author{H.~T.~Nguyen\altaffilmark{5}}
\author{N.~Ponthieu\altaffilmark{12}}
\author{C.~Pryke\altaffilmark{11}}
\author{S.~Richter\altaffilmark{3}}
\author{G.~Rocha\altaffilmark{3,5}}
\author{K.~W.~Yoon\altaffilmark{13}}

\altaffiltext{1}{Physics Department, University of California, Berkeley, CA 94720, USA}
\altaffiltext{2}{University of Wales, Cardiff, CF24 3YB, Wales, UK}
\altaffiltext{3}{California Institute of Technology, Pasadena, CA 91125, USA}
\altaffiltext{4}{National Radio Astronomy Observatory, Santiago, Chile}
\altaffiltext{5}{Jet Propulsion Laboratory, Pasadena, CA 91109, USA}
\altaffiltext{6}{University of California, San Diego, CA 92093, USA}
\altaffiltext{7}{Princeton University, Princeton, NJ 08544, USA}
\altaffiltext{8}{Commissariat \`a l'Energie Atomique, Grenoble, France}
\altaffiltext{9}{Institut d'Astrophysique de Paris, Paris, France}
\altaffiltext{10}{Stanford University, Palo Alto, CA 94305, USA}
\altaffiltext{11}{University of Chicago, IL 60637, USA}
\altaffiltext{12}{Institut d'Astrophysique Spatiale, Universit\'e Paris-Sud, Orsay, France}
\altaffiltext{13}{National Institute of Standards \& Technology, Boulder, CO 80305, USA}

\begin{abstract}
The \bicep\ experiment was designed specifically to search for the
signature of inflationary gravitational waves in the polarization of the 
cosmic microwave background (CMB).
Using a novel small-aperture refractor and 49 pairs of 
polarization-sensitive bolometers, \bicep\ has completed 3 years of 
successful observations at the South Pole beginning in 2006 February. 
To constrain the amplitude of the inflationary $B$-mode polarization, 
which is expected to be at least 7 orders of magnitude fainter than 
the 3 K CMB intensity, precise control of systematic effects is essential.
This paper describes the characterization of potential systematic errors for 
the \bicep\ experiment, supplementing a companion paper on 
the initial cosmological results. 
Using the analysis pipelines for the experiment, we have simulated the impact of 
systematic errors on the $B$-mode polarization measurement. 
Guided by these simulations, we have established benchmarks for the characterization
of critical instrumental properties including bolometer relative gains, 
beam mismatch, polarization orientation, telescope pointing, sidelobes, 
thermal stability, and timestream noise model.
A comparison of the benchmarks with the measured values shows that we have 
characterized the instrument adequately to ensure that systematic errors 
do not limit \bicep's 2-year results,
and identifies which future refinements are likely necessary to probe
inflationary $B$-mode polarization down to levels below a tensor-to-scalar 
ratio $r = 0.1$.
\end{abstract}


\section{Introduction}
\setcounter{footnote}{0}

A strong indication of an inflationary origin of the universe would be 
a detection of the curl component (``$B$-mode'') in the polarization of 
the CMB arising from gravitational wave perturbations \citep{Dodelson2009}.
This primordial $B$-mode polarization is expected to peak at angular 
scales of $\sim$2$\deg$, and the magnitude of the power spectrum is 
described by the ratio $r$ of the initial tensor-to-scalar perturbation 
amplitudes, a quantity directly related to the energy scale of inflation.  
The best published upper limit is $r<0.22$ at 95\% confidence, derived 
from $WMAP$ CMB temperature anisotropy measurements at large angular scales 
combined with constraints from Type Ia supernovae and baryon 
acoustic oscillations \citep{Komatsu2009}.
Upper limits on the $B$-mode polarization amplitude, 
$\sqrt{\ell(\ell+1)C_\ell^{BB}/2\pi}$, of $\sim$0.8 $\mu$K rms have been 
placed by $WMAP$ at a multipole moment of $\ell$$\sim$65 and by \QUAD\ at 
$\ell$$\sim$200, respectively \citep{Nolta2009,Pryke2009}.
These limits are still well above the expected levels of confusion from 
either polarized Galactic foregrounds in the cleanest regions of the sky 
or from gravitational lensing that converts the much brighter
CMB gradient (``$E$-mode'') polarization to $B$-modes at smaller angular 
scales.

\bicep\ (Background Imaging of Cosmic Extragalactic Polarization) is 
an instrument designed to target the expected peak of the 
gravitational-wave signature at angular scales around 2$\deg$.
Using proven bolometric technologies and selecting the cleanest available 
field for observation, this instrument 
was designed, given sufficient observation time,
to be capable of measuring a polarization signal of 
0.08~$\mu$K rms at $\ell$$\sim$100 corresponding to the BB signal
expected for a tensor-to-scalar ratio of $r=0.1$. 
The CMB temperature anisotropy has a much larger amplitude of $\sim$50~$\mu$K 
rms at these angular scales, and imperfect rejection of it in 
the polarization measurement could result in a residual false signal.  
In addition, errors in polarization orientations and pointing could mix 
the $\sim$1~$\mu$K $E$-mode CMB polarization signal into spurious 
$B$-modes.
This experiment therefore requires careful instrument characterization and 
calibration to minimize systematic contamination in the polarization 
measurement.

This paper supplements a companion paper that presents the CMB 
polarization power spectra from the first 2 years of \bicep\ data 
\citep{Chiang2010}.
The rest of this section gives an overview of the instrument, the
observing strategy, and the collected data.  Section 2 describes  
the simulations that determine the impact of the instrumental parameter 
characterization on the cosmological results.  
Section 3 describes each of the instrumental properties and how we 
calibrate them.
Section 4 discusses how we characterize the properties of noise in our 
data and quantify the noise bias through simulations.
We conclude by identifying what we find to be the most important 
systematic uncertainties for \bicep\ and discussing the path forward
toward improving these uncertainties for future, more sensitive, measurements.

\subsection{Instrument design overview}

\begin{figure*}[tb]
\begin{center} \begin{tabular}{c}
\resizebox{\textwidth}{!}{\includegraphics{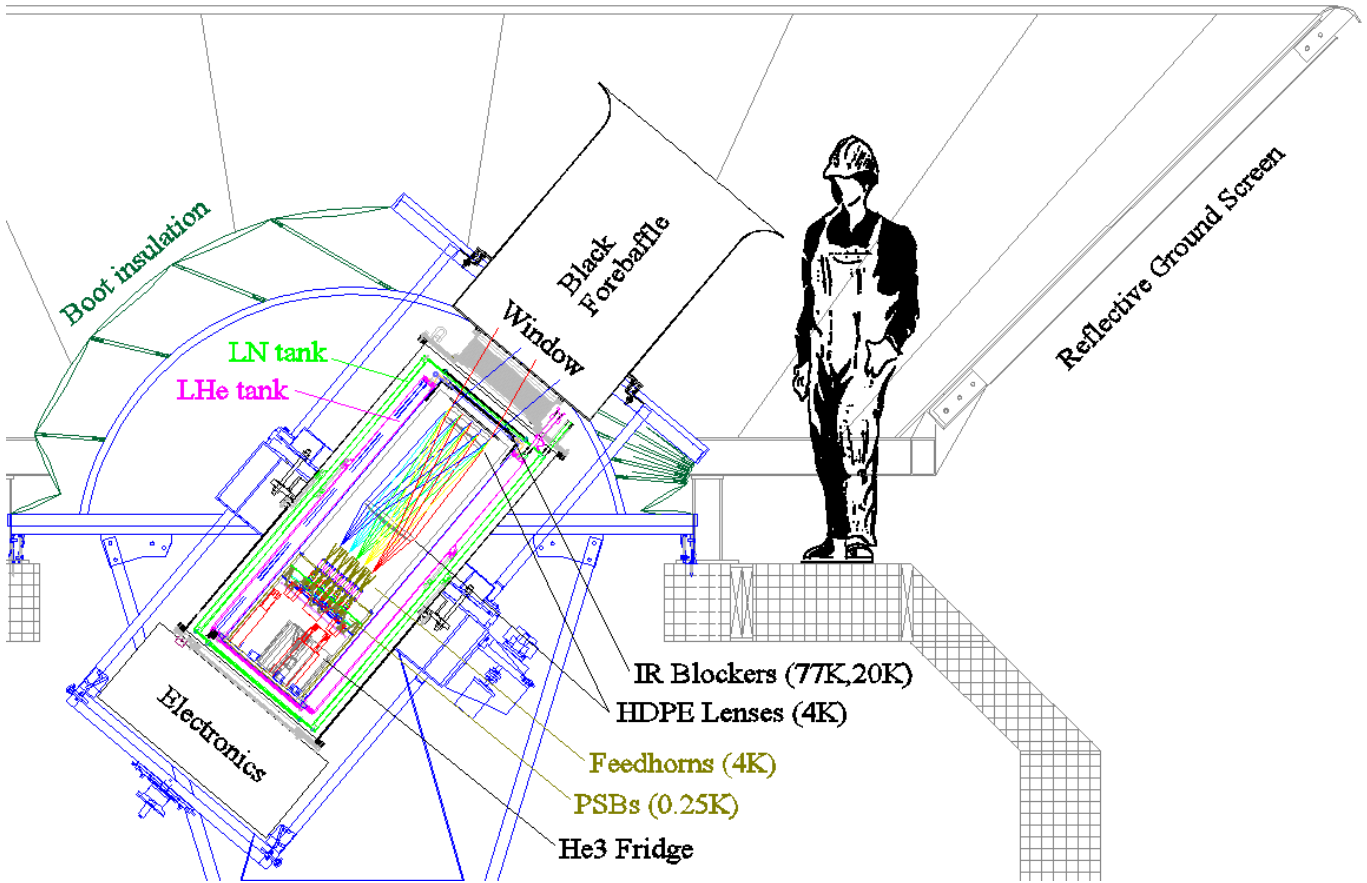}} 
\end{tabular} \end{center}
\caption{\label{fig:instrument} \bicep\ telescope on a 3-axis mount at its 
lowest elevation limit of 50$\deg$, looking out through the roof of 
the Dark Sector Laboratory (DSL) building located 800~m from the 
geographic South Pole. 
A cryostat with toroidal liquid nitrogen and liquid helium tanks encloses 
the entire 4~K optics, including two high-density polyethylene lenses and 
corrugated feedhorns.  The polarization-sensitive bolometers are cooled 
with a $^4$He/$^3$He/$^3$He sorption refrigerator to 250~mK. }
\end{figure*}

The goal of targeting the sub-$\mu$K $B$-mode polarization signal that 
peaks at $\sim$2$\deg$ angular scales led to an experiment design 
optimized especially for sensitivity and control of systematic errors.
The design of \bicep\ and the observation strategy are described in 
\citet{Yoon2006} and \citet{Keating2003a}. 
\bicep\ (Figure~\ref{fig:instrument}) is a compact on-axis refractor with 
49 pairs of polarization-sensitive bolometers \citep[PSBs;][]{Jones2003}
operating in atmospheric transmission windows near the CMB peak at 100 
and 150~GHz with $0.9\deg$ and $0.6\deg$ beams, respectively 
(Table~\ref{tab:summary}).
We observe in two frequency bands to differentiate between the spectra of CMB 
anisotropies and sources of potential Galactic foreground contamination.
The Amundsen-Scott South Pole Station, at an elevation of 2800~m, was 
chosen for its atmospheric 
transparency, stable weather, and constant availability of an excellent
observing field on the sky.
Achieving one degree resolution at 2--3~mm wavelengths requires only a 25~cm 
aperture, which is compatible with a compact forebaffle and simple implementation of 
calibration measurements.

The bolometers use neutron transmutation doped (NTD) germanium thermistors 
to measure the optical power incident on a polarization-sensitive
absorber mesh.  
After adjusting for relative responsivities, orthogonal PSBs within a pair 
are summed or differenced to obtain temperature or polarization measurements.  
Because the orthogonal PSBs observe the CMB through the same optical path and 
atmospheric column with nearly identical spectral passbands, systematic 
contributions to the polarization are minimized.

\begin{table}[!ht]
\caption{\label{tab:summary} \bicep\ Instrument Summary}
\centering \begin{tabular}[c]{ccccc} \hline\hline
Band Center & Bandwidth & Beam FWHM$^a$ & PSBs & NET$^b$ Per Detector \\
\hline
~~96.0 GHz & 22.3 GHz & 0.93$\deg$ & 50 & 530 $\mu$K$_{\rm CMB}\sqrt{s}$ \\
150.1 GHz & 39.4 GHz & 0.60$\deg$ &~~48$^c$ & 450 $\mu$K$_{\rm CMB}\sqrt{s}$ \\
\hline
\multicolumn{5}{l}{\scriptsize $^a$ Full width at half maximum, average over all the beams.} \\
\multicolumn{5}{l}{\scriptsize $^b$ Noise equivalent temperature; see \S\ref{sec:noise}.} \\
\multicolumn{5}{l}{\scriptsize $^c$ After the first year, two of the 150~GHz pairs were converted to 220~GHz.}
\end{tabular}
\end{table}

\begin{figure}[!ht]
\begin{center}\begin{tabular}{c}
\includegraphics[trim=0 0 0 35,clip,width=\linewidth]{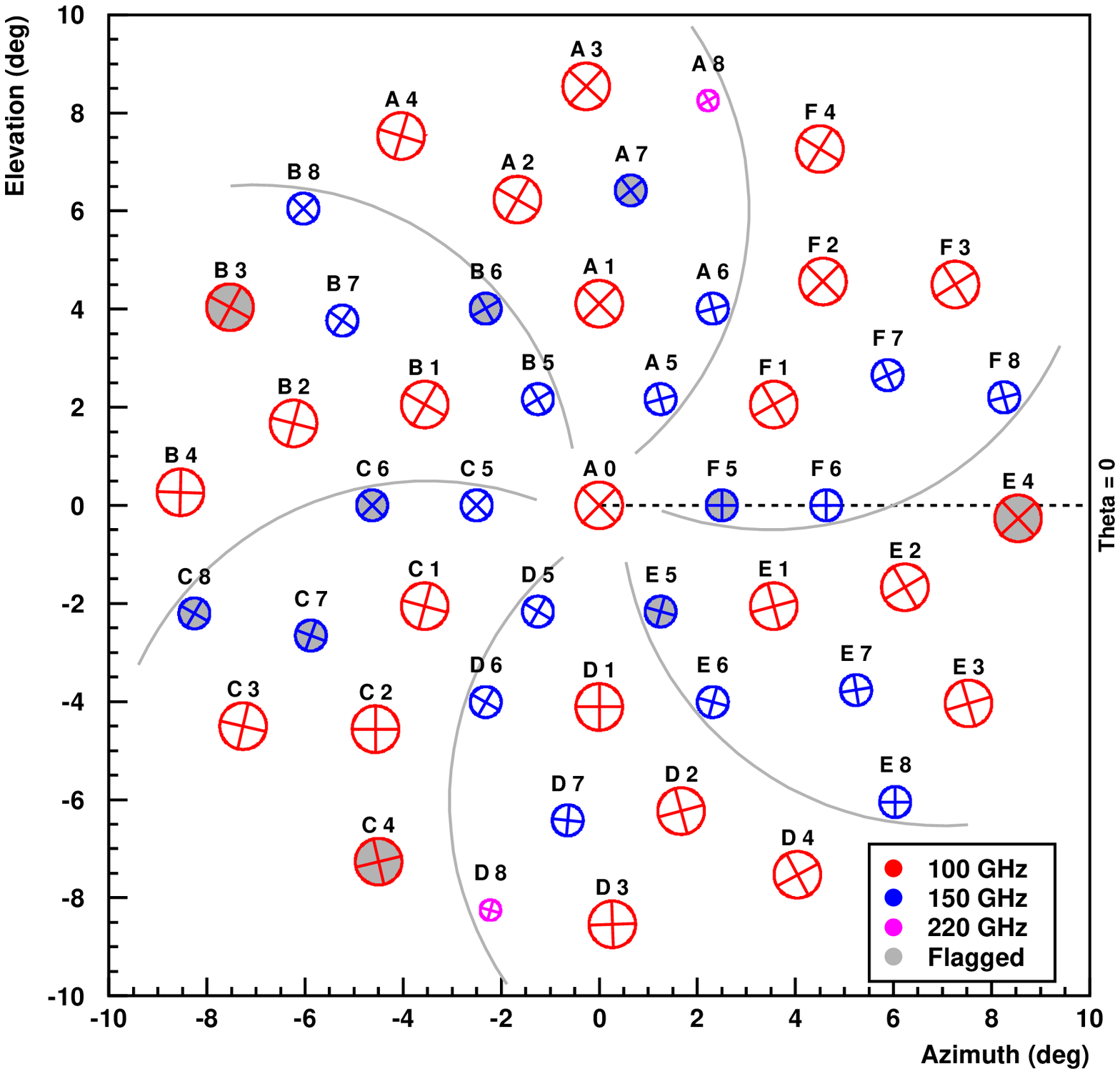}
\end{tabular}
\end{center}
   \caption[layout]
   {\label{fig:layout}
2007/2008 layout of the \bicep\ beams with nominal locations, FWHM, and 
polarization orientations.  
There are six sections (separated by curved gray lines) having alternating 
``Q'' or ``U'' PSB orientations with respect to the center.  
The observations are performed with the focal plane orientations of 
$-45\deg, 0\deg, 135\deg,$ and $180\deg$ counterclockwise about the boresight, 
providing two independent and complete polarization coverages of the field 
to allow jackknife tests.
(A 180$\deg$ boresight rotation provides complete coverage of $Q$ and $U$ 
Stokes parameters on the sky at each beam location.) }
\end{figure}

The PSB layout on the focal plane (Figure~\ref{fig:layout})
was chosen so that a 180$\deg$ rotation about the boresight completely 
exchanges the polarization coverage on the sky.
At the end of the first year, in 2006 November, we added prototype 220~GHz 
feedhorns in place of two of the 150~GHz ones along with the appropriate 
filters.  
We also replaced four bolometers because of their slow temporal response, 
high noise level, or poor polarization efficiency.
We have omitted these and other problematic PSB pairs from CMB analysis
for each observing year.
After this refurbishment, \bicep\ remained cold and operated without
interruption until the 
completion of the observations in December 2008.

\subsection{Observing strategy} \label{sec:strategy}

With an instantaneous field of view spanning 18$\deg$, \bicep\ maps an 
800~deg$^2$ field daily by scanning the boresight in azimuth over a 
64$\deg$ range at 2.8$\deg$/s with hourly $0.25\deg$ steps in elevation 
from 55$\deg$ to 60$\deg$.  At each elevation step, the telescope 
completes 50 back-and-forth azimuth scans, or 100 ``half-scans,'' making 
up a ``scan set.'' The scan speed was selected so that our target angular 
scales ($\ell\sim$~30--300) appear at 0.1--1~Hz, above a significant 
portion of the $1/f$ atmospheric noise, while limiting motion-induced 
thermal fluctuations at the detectors.

\begin{figure*}[!htb]
\begin{center} \begin{tabular}{c}
\includegraphics[trim=15 5 5 5,clip,width=\linewidth]{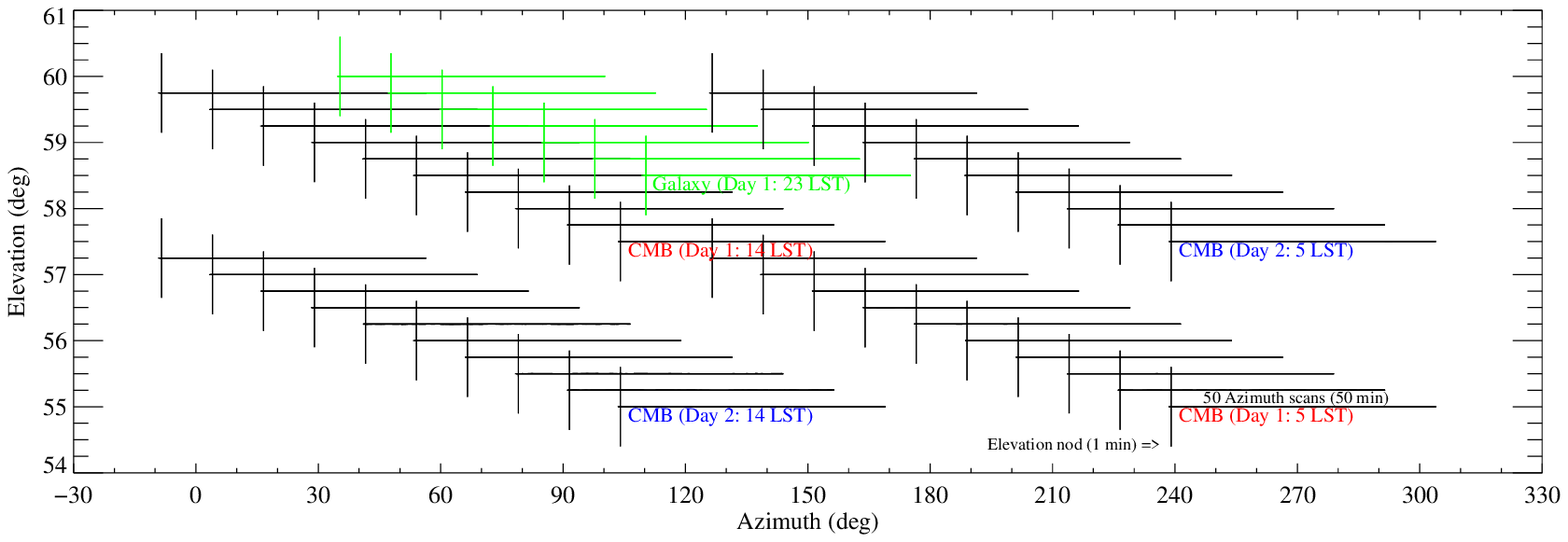}
\end{tabular} \end{center}
\caption[cycle]{\label{fig:cycle} A 48-hr observing cycle consists of 
6 hours for cycling the refrigerator (not shown), two 9-hr blocks of 
raster scans on our main CMB field on the first day, a 
6-hr block of similar raster scans on the Galaxy, and two more 9-hr 
blocks on the main CMB field on the second day.
Each of the upper and lower halves of the elevation range is scanned in 
different azimuth ranges between the two days, 
allowing a jackknife test for ground contamination.  
At the beginning and end of each set of 50-minute azimuth scans, a 
$\pm0.6\deg$ elevation ``nod'' is performed to measure relative gains of 
every bolometer.
This 48-hr cycle is repeated at different boresight rotation angles: 
$\{-45\deg, 0\deg, 135\deg, 180\deg\}$. }
\end{figure*}

\bicep\ operates in a 48 sidereal-hour observing cycle 
(Figure~\ref{fig:cycle}), with each cycle at one of the four fixed 
boresight rotation angles $\{-45\deg, 0\deg, 135\deg, 180\deg\}$.
A $45\deg$ rotation about the boresight exchanges the polarization
coverages on the sky, providing two independent pairings of angles, 
each with complete polarization coverage per overlapped sky pixel.
The two sets rotated by 180$\deg$ also help to average down systematic 
effects like differential pointing.

Each 48-hr cycle begins with 6 hours allocated for recycling the
refrigerator, filling liquid nitrogen (every 2 days) and liquid helium
(every 4 days), and performing optical star pointing calibrations
along with a mount tilt measurement. 
The CMB field is completely mapped once each day in two 9-hr blocks, 
with the scan order of the upper and lower halves of the elevation range 
switched such that the azimuth ranges for the two days are offset.
Differencing the first and second days of a given 48-hr observing cycle 
tests for potential azimuth-fixed contamination. 
Overlapping coverage of the sky from detectors with various polarization 
orientations is created by scanning in azimuth, stepping in elevation, and 
rotating the telescope with respect to the boresight every 2 days.

Each scan set at a given elevation is fixed with respect to a given 
azimuth range instead of tracking the field center over the hour-long period. 
This scan strategy allows for a straightforward removal of any azimuth- or 
scan-synchronous contamination.  For each scan set and each of the two 
scan directions, the entire timestream is simply binned in azimuth to form 
a template signal which is subtracted from each half-scan.

Relative detector responsivities are measured at the beginning and end of 
each 1-hr fixed-elevation scan set by fitting the detector response to 
a small change in line-of-sight airmass (``elevation nods''), described 
in \S~\ref{sec:elnod}.

\subsection{Collected data and observing efficiency}

The \bicep\ mount was installed at the South Pole in 2005 November, the 
cryostat was first cooled in December, and the instrument
captured first astronomical light a month later.  
Following calibration measurements and 
tests of the observing strategy, \bicep\ began CMB observations in 2006
February.  The instrument operated nearly continuously until 2008 
December, when it was decommissioned to prepare for its   
replacement by \bicep2 on the same mount, planned for late 2009.

Excluding any incomplete 9-hr observing blocks, \bicep\ acquired 180 
days of CMB observations during 2006, in which a significant fraction of 
the observing season was devoted to calibration measurements.  The amount 
of CMB observations increased to 245 days in 2007 and a similar amount in 
2008.
Although there is no evidence for Sun contamination during the summer, we 
restrict our CMB analysis to data taken during February--November.

The first 2.5 months of data in 2006 were excluded from the current 
analysis because a different scan strategy was being investigated at this 
time.
As a coarse weather cut, we have excluded 9-hr blocks if the relative 
gains derived from elevation nods vary by more than 20\% rms,
averaged over the channels.  
This criterion cuts clear outliers in the distribution of atmospheric 
instability.
After these cuts, 117 days in 2006 and 226 days in 2007 remain for our 
baseline CMB analysis (Table~\ref{tab:observation}).

Furthermore, 3\% of PSB pair timestreams are omitted due to cosmic ray 
hits, glitches, or $>$3\% mismatch in the relative gain measured at the 
beginning and end of each 1-hr scan set.  
Accounting for the 75\% scan efficiency and the scheduled 
calibration routines, the net CMB observing efficiency is 60\% during the 
CMB observing blocks and 45\% overall during each 2-day cycle.

\begin{table}[!ht]
\caption{\label{tab:observation} CMB Observation Summary}
\centering \begin{tabular}[c]{lcccc} \hline\hline
Year & \multicolumn{2}{c}{PSB Pairs: used (total)} & Observing Days: & Integration \\
     & 100 GHz & 150 GHz & used (total) & Time$^a$ \\
\hline
2006 & 19 (25) & 14 (24) & 117 (180) days & 4.5$\times10^6$ s \\
2007 & 22 (25) & 15 (22) & 226 (245) days & 8.8$\times10^6$ s \\
\hline
\multicolumn{5}{l}{\scriptsize $^a$ Based on 18 hours per day of CMB 
observation at 60\% net observing efficiency.}
\end{tabular} 
\end{table}

\section{Calibration goals and systematic error simulation}\label{s:simulation}

To process the timestreams into co-added polarization maps with 
systematic errors tolerable for our target sensitivity, an accurate 
characterization of the detectors and their beams is essential.
Imperfections in the experiment and its characterization can result in 
false $B$-mode polarization signal.
Many of these systematic effects depend in a complex
way on the scan strategy, so analytic estimates of
the impact of an instrumental uncertainty on the final power spectra
serve only as a rough guide.
Using the actual analysis pipelines, we have simulated the most significant 
instrumental uncertainties to establish benchmarks for how precisely each 
property must be measured.
The results are summarized in Table~\ref{tab:systematics}.
Calibration uncertainties that affect only the power spectrum amplitudes,
but do not cause false polarization signals, are summarized in 
Table~\ref{tab:calibration} and discussed in the last paragraph of this section.

We are most concerned with systematic errors that mix temperature 
anisotropy $T$ and $E$-mode polarization signals into $B$-mode 
polarization.
Instrumental properties that must be characterized are discussed in detail 
in the next section.
The response of a PSB to radiation characterized by Stokes \{$T,Q,U$\}, 
which are functions of frequency $\nu$ and direction $\Omega$, can be 
modeled as:
\begin{eqnarray} \label{eq:model}
d(t) = K_t * {\Big\{}n(t) + & &		
g \int d\nu A_e F(\nu) \int d\Omega\; P(\Omega) \nonumber \\ 
& & [ T + \frac{1-\epsilon}{1+\epsilon} (Q \cos{2\psi} + U \sin{2\psi}) ] {\Big\}},
\end{eqnarray}
where $\psi$ is the polarization orientation angle of the PSB, $\epsilon$ 
is the cross-polarization response, $P(\Omega)$ is the beam function, 
$F(\nu)$ is the spectral response, $A_e$ is the effective antenna area, 
$g$ is the responsivity at 0 Hz, $n(t)$ is noise, and $K_t$ is the 
time-domain impulse response associated with the detector's frequency 
transfer function.
The temporal response function of each bolometer must be measured 
to deconvolve it from the raw timestream.  Then the relative gain within 
each PSB pair must be determined for differencing.  Since we derive relative 
gains from the atmospheric signal in elevation nods, we must verify that 
the spectral response of each PSB pair is well matched.  In addition, the 
polarization differencing requires that the two PSBs have well 
matched beam shapes and pointing.  
Finally, construction of the polarization map requires the knowledge of 
the polarization orientations of the PSBs and the telescope pointing.

We established our calibration benchmarks based on \bicep's design goal to
be capable of measuring polarization down to levels corresponding to 
a tensor-to-scalar ratio $r=0.1$ without being limited by systematic effects.  
At $r=0.1$, the $B$-mode polarization power spectrum, $C_\ell^{BB}$, would 
have a peak amplitude of $\ell(\ell+1)C_\ell^{BB}/2\pi$ = 0.007~$\mu$K$^2$ 
at $\ell$ = 70--110.  
The benchmarks for the instrumental properties and characterization 
correspond to the values that result in spurious $B$-mode signal at the 
level of $r=0.1$ in the simulations.

The simulation procedure uses the same data processing pipelines as the 
main CMB power spectrum analysis, and is basically identical to the 
signal-only simulations used in determining the $\ell$-space filter 
function.
We verified the simulation results using our two 
independent pipelines: one using \QUAD's pseudo-$C_\ell$ estimator on 
a flat sky, and the other using \spice\ \citep{Chon2004} on a curved sky.

We begin with a \lcdm\ model generated by \camb~\citep{Lewis2000}, using 
cosmological parameters derived from $WMAP$ 5-year 
data~\citep{Hinshaw2009} and $r = 0$.  From this model, we generate an 
ensemble of simulated CMB skies using \synfast~\citep{Gorski2005}.  
We then simulate observations of the \bicep\ field on these \synfast\ 
skies using pointing data from the actual scan patterns.
We vary instrumental parameters for the PSB pairs with a distribution 
across the array corresponding to the given uncertainty, but constant in time.
We have assumed random distributions, although the differential beams 
could have a pattern across the focal plane resulting in 
more or less false signal.
For differential pointing, we use the actual measured quantities in the 
simulations.

\begin{table*}[!hbt]
\caption{\label{tab:systematics}
\small Systematic errors potentially producing false $B$-mode polarization}
\centering \begin{tabular}[c]{lcclc} \hline\hline
Instrument Property & Benchmark$^a$   & Measured & Measurement notes & Reference \\
\hline
Relative gain uncertainty: $\Delta(g_1/g_2)/(g_1/g_2)$& 0.9\%	& $<1.1\%$
& Upper limit, rms error over the array.$^b$ & \S\ref{sec:relgains} \\
Differential pointing: ~~$(~{\bf r_1}-{\bf r_2}~)/\sigma$ $^c$	& 1.9\%&~~~~1.3\%
& Average, each repeatedly characterized to 0.4\% precision.$^d$ & \S\ref{sec:beams} \\
Differential beam size: $(\sigma_1-\sigma_2)/\sigma$ & 3.6\% 	& $<0.3$\%
& Upper limit, rms over the array. & \S\ref{sec:beams} \\
Differential ellipticity: $(~e_1-e_2~)/2$	& 1.5\%		& $<0.2$\%
& Upper limit, rms over the array. & \S\ref{sec:beams} \\
Polarization orientation uncertainty: $\Delta\psi$ & 2.3$\deg$	& $<0.7\deg$
& Upper limit, rms absolute orientation error over the array. & \S\ref{sec:polarization} \\
Telescope pointing uncertainty: $\Delta{\bf b}$	& 5$^{\prime}$	& 0.2$^{\prime}$
& Fit residual rms in optical star pointing calibration. & \S\ref{sec:pointing} \\
Polarized sidelobes (100, 150 GHz)		& -9, -4 dBi	& -26, -17 dBi
& Response at 30$\deg$ from the beam center. & \S\ref{sec:sidelobe} \\
Focal plane temperature stability: $\Delta$$T_{\rm FP}$ & 3 nK	& 1 nK
& Scan-synchronous rms fluctuation on $\ell$$\sim$100 time scale. & \S\ref{sec:thermal} \\
Optics temperature stability: $\Delta$$T_{\rm RJ}$ & 4 $\mu$K  & 0.7 $\mu$K
& Scan-synchronous rms fluctuation on $\ell$$\sim$100 time scale. & \S\ref{sec:thermal} \\
\hline
\multicolumn{5}{l}{\scriptsize $^a$ Benchmarks correspond to values that result in
a false $B$-mode signal of at most $r=0.1$.  For $r=0.01$, all benchmarks would be lower by $\sqrt{10}$.} \\
\multicolumn{5}{l}{\scriptsize $^b$ If relative gain errors are detected, 
we anticipate removing their effects in future analyses using a CMB temperature template map.} \\
\multicolumn{5}{l}{\scriptsize $^c$ $\sigma = FWHM/\sqrt{8\ln(2)}$ = 
\{0.39$\deg$, 0.26$\deg$\} at \{100, 150\} GHz.} \\
\multicolumn{5}{l}{\scriptsize $^d$ This measurement of differential pointing could be used
in future analyses to remove the small predicted leakage of CMB temperature into polarization maps.} \\
\end{tabular} 
\end{table*}

To simulate the coupling between non-ideal beams and the sky, we follow 
the formalism in \citet{Bock2008}.
A PSB timestream sample is expressed as a convolution of the beam with a 
second-order Taylor expansion of the sky signal around the pointing center---the 
first and second derivatives of simulated $T$, $Q$, and $U$ maps 
are calculated with \synfast.  
Using this technique, the beam convolution 
can be simulated quickly while using the exact scan trajectory of each 
detector, which is essential in quantifying the beam mismatch 
effects, as these depend on the scan strategy and on the combination of 
detectors with different characteristics.
This formalism also allows simulation of the impact of uncertainties in 
the knowledge of PSB orientations and cross-polarization response.

The simulated observation is performed at each of the 4 boresight rotation 
angles and with the appropriate instrument configuration for each 
observing year.  
The simulated signal-only timestreams are fed through the pipelines to be 
filtered and co-added into maps with exactly the same weights as with the 
real data.  
A baseline set of spectra are computed without any simulated errors to (1) 
determine the amount of $E$-$B$ leakage due to timestream filtering
and any other effects intrinsic to the pipeline, and (2) obtain the 
transfer functions to be applied to the raw spectra.

The power spectra of maps made with simulated errors are compared with 
those without any errors, and any differences---in particular, excess 
$B$-mode polarization power that has leaked from $T$ or $E$---are 
attributed to the systematic errors.
The simulations are performed with at least 10 realizations of input CMB 
maps.

Table~\ref{tab:systematics} summarizes the instrument properties and
$r=0.1$ benchmark levels for their characterization,
as well as the results of measurements described in the next section.
Each instrumental property has been characterized to a level of precision 
at least comparable to the $r=0.1$ benchmark and in most cases much better.

Finally, some calibration uncertainties affect only the scaling 
of the spectra but do not result in spurious polarization signals
(Table~\ref{tab:calibration}).  
These effects can be calculated analytically, and are described in the 
next section.
We set the benchmark for each of these uncertainties such that the contribution
to the calibration of power spectrum amplitudes is $\Delta C_\ell / C_\ell 
\leq 10\%$, a standard easily met by our instrument characterization.

\begin{table}[!ht]
\caption{\label{tab:calibration} Calibration uncertainties \\
affecting the power spectrum amplitudes only}
\centering \begin{tabular}[c]{lcc} \hline\hline
Calibration Quantity & Benchmark$^a$ & Measured \\
\hline
Absolute gain $\Delta g/g$		&	$5\%$ 	& $2\%$ \\
Cross-polarization response $\Delta\epsilon$ &	$0.026$ & $0.01$ \\
Relative polarization orientation $\Delta(\psi_1-\psi_2)$ & $9\deg$ & $0.1\deg$ \\
\hline
\multicolumn{3}{l}{\scriptsize $^a$ Benchmarks correspond to 10\% 
uncertainty in the polarization power spectrum} \\
\multicolumn{3}{l}{\scriptsize amplitude.}
\end{tabular} 
\end{table}

\section{Instrument characterization} \label{sec:characterization}

We use the benchmarks found in the previous section to guide our program to 
characterize the instrumental performance. 
To reach the required level of precision, we designed and implemented a 
number of techniques for the calibration measurements.
This section describes the measurements for relative gain calibrations, 
beam characterization, polarization calibration, telescope pointing, far 
sidelobes, and thermal stability and compares the results with the 
benchmark values leading to a reference level of uncertainty in $BB$ power 
spectrum.  
The characterization of the instrument noise and its implications for
uncertainty in the power spectrum is discussed in 
\S\ref{sec:noise}.  All the instrumental quantities have been 
measured with sufficient accuracy for the sensitivity achieved with the 
first 2 years of data.

\subsection{Relative detector gains} \label{sec:relgains}

Polarization measurement with \bicep\ relies on PSB pair differencing.  
The relative gains within each pair must be accurately determined 
to prevent unpolarized signal from leaking into the polarization measurement.
The relative amplitudes of the CMB temperature and polarization 
anisotropies place 
stringent requirements on the relative gain calibration of the differenced detectors.
Simulations of relative gain errors suggest 
that the level of false polarization depends strongly on how gain errors 
are distributed across the PSB array.
With the best estimated distribution of the measured gain uncertainties 
described in \S\ref{sec:elnod}, the simulations indicate that relative 
gains need to be accurate to 0.9\% rms to limit the leakage of CMB 
temperature anisotropy into $B$-mode power at a level corresponding to 
$r=0.1$.

We correct for relative gain differences in two steps, first by 
deconvolving the temporal transfer function of each PSB to account for 
frequency-dependent gains, and then by correcting for the DC gains through 
elevation nods.
Since elevation nods use the 
atmospheric emission, which has a different emission spectrum than the CMB 
temperature fluctuations, the spectral response of each PSB in a pair must 
be precisely matched.
We measured the detector transfer functions each year, calibrated 
the DC gains hourly with elevation nods, and made spectral response 
measurements once before deployment and once in the field.
The following subsections describe the measurements of temporal transfer function, 
DC responsivities, and spectral response.
The transfer functions for each pair were measured to within 0.3\% 
uncertainty and elevation nod relative gains to within 1.1\% rms 
uncertainty over the PSB pairs.  
While the current upper limit on the possible relative 
gain errors slightly exceeds the 0.9\% benchmark for $r=0.1$, this effect 
is not significant for the 2-year CMB results; furthermore, if a significant
signal was detected it could be corrected with a more sophisticated analysis.

\subsubsection{Temporal transfer functions}

Analysis of the time series from each detector begins by deconvolving the 
temporal response using the measured frequency-domain optical transfer 
function of the detector.  Since the transfer function is proportional to 
the gain of the detector as a function of frequency, it directly affects the 
relative gains of a PSB pair to be differenced.  The relative transfer 
functions must thus be measured with errors below the 0.9\% benchmark 
set for the relative gains.

At the nominal scan speed of $2.8\deg$/s in azimuth at $\sim$60$\deg$ 
elevations, our target angular scales of $\ell=$~30--300 fall 
into the frequency band of approximately 0.1--1~Hz.  Since the elevation 
nods described in the following 
section are sensitive to relative fluctuations 
at $\sim$0.02~Hz, the transfer functions were measured down to 0.01~Hz.

\begin{figure}[!htb]
\includegraphics[width=\linewidth]{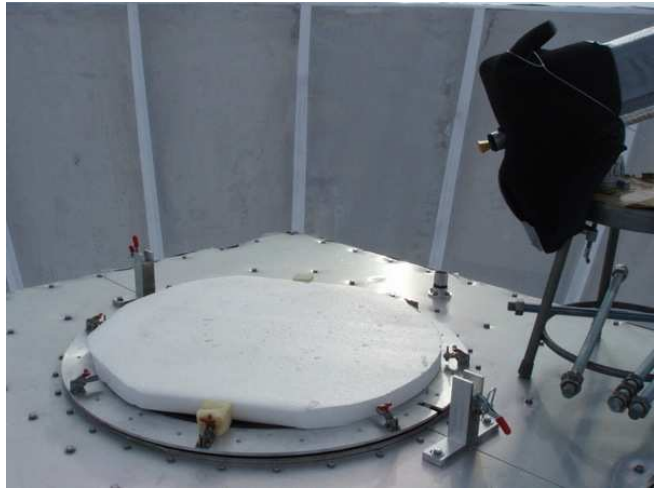}
\caption[scatter] 
{\label{fig:scatter} Modulated source in front of the telescope aperture used 
for the measurement of transfer functions.  Metal washers are embedded in 
transparent polypropylene foam sheet to scatter the PIN-switched broadband 
noise source signal (from upper right) into the beam while keeping the 
total loading similar to that during nominal CMB observations.  }
\end{figure}

\begin{figure}[!htb]
\includegraphics[width=\linewidth]{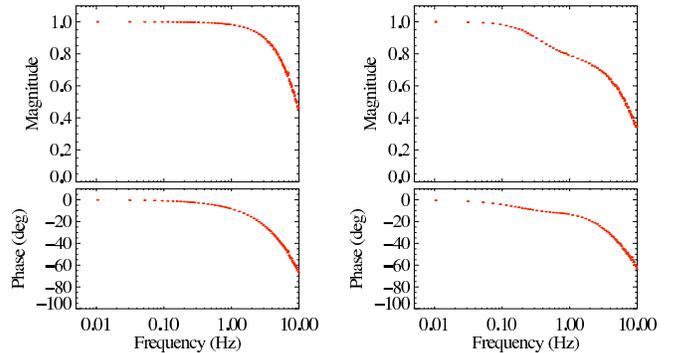}
\caption[transfer]
{\label{fig:transfer} Measured transfer functions for a representative 
PSB (left) and an anomalous PSB (right), with error bars showing rms 
measurement repeatability. 
Most of the anomalous transfer functions are well described by 
Equation~\ref{eq:transfer} and are repeatably measured, but are 
conservatively excluded from the initial analysis.}
\end{figure}

The primary measurement technique involved analyzing the step response to 
a fast-switched square-wave source (Gunn oscillator or broadband noise 
source) operating at 0.01~Hz, while under optical loading conditions 
representative of CMB observations (Figure~\ref{fig:scatter}).
Possible dependence on background loading and detector non-linearity were 
explored by repeating the measurement with extra loading from sheets of
emissive foam placed in the beam in combination with different signal 
strengths.
The ratio of the Fourier transforms of the time-domain detector response 
and of the input square wave were averaged for each detector to obtain the 
transfer function.
The measured transfer functions for a representative and anomalous PSBs 
are shown in Figure~\ref{fig:transfer}; they were stable against different 
combinations of loading and signal levels.

The relative gain uncertainty due to measurement uncertainty is found to 
be $<$0.3\% over the frequency range of 0.01--1~Hz.
The measured transfer functions fit the following model as a function of 
frequency $\omega$,
\begin{equation} \label{eq:transfer}
{\tilde K}(\omega) \propto 
\frac{1-\alpha}{(1-i\omega\tau_1)(1-i\omega\tau_2)}
+ \frac{\alpha}{(1-i\omega\tau_\alpha)},
\end{equation}
where $\tau_{1,2,\alpha}$ are time constants and $\alpha$ is the 
fractional amount of a slow additive component.
However, we measured the transfer functions with sufficiently high 
signal-to-noise ratio so that they could be directly inverse Fourier 
transformed to define the deconvolution kernels.
The median time constants were $\tau_1 \sim 20$~ms and $\tau_2 \sim 5$~ms, 
and $\tau_\alpha$ is generally 100--200~ms with $\alpha$ typically 
$<$ 0.05.
From the first observing year, 6 channels at 150~GHz were excluded from 
CMB analysis due to excessive roll off between 0.01 and 0.1~Hz (large 
$\alpha$ and $\tau_\alpha$). 
Two of the worst were in a single PSB pair and were replaced at the end of the year.

Between the first two observing years, the transfer function measurements 
generally agreed to within 0.5\% rms across the signal band.  Two 
exceptions were excluded from the first year CMB data where the transfer 
functions were less well constrained at low frequencies.
Details of the measurements and analysis are in \citet{Yoon2007}.

\subsubsection{Relative responsivities} \label{sec:elnod}

Relative gains are derived from elevation nods performed at the 
beginning and end of every 1-hr constant-elevation scan 
set (Figure~\ref{fig:cycle}).  The telescope scans in elevation with 
a rounded triangle wave pattern over a range of $\pm0.6\deg$, 
varying the optical loading by $\sim\pm$0.1~K due to the changing 
line-of-sight air mass. 
The bolometer responses are fit to a simple air mass model of atmospheric 
loading versus elevation, $T_{atm} \propto$ csc($EL$), to derive the 
relative gains across the array.

The elevation nod is performed slowly over 50~s to limit 
thermal disturbances on the focal plane.
During the nod, the diagnostic ``dark'' PSBs not illuminated 
through the feedhorns exhibit systematic voltage responses
(also seen in the focal plane thermistors but not in the resistor channels)
that are 
$\sim$0.4\% of the typical responses of the illuminated PSBs, indicating 
thermal contamination at this level.  
To reduce the effect of the thermally-induced signals, the two 
elevation nods for each scan set are performed in opposite patterns 
(up-down-return and down-up-return) and the average response is used.  
While the two patterns result in a small systematic difference in the 
individual gains, the PSB pair {\it relative} gains are consistent to 
within the measurement noise of 0.3\% rms.

\begin{figure}[!htb]
   \centering
   \includegraphics[width=\linewidth]{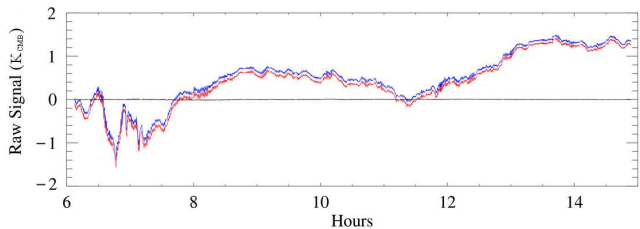}
   \caption[diff]
   { \label{fig:diff} Individual 150 GHz timestreams within a PSB pair 
(red and blue) are differenced (black) in this plot using a single 
relative gain fit over the plotted 9-hr period.  
For the actual CMB analysis, relative gains are updated for every 1-hr 
scan set.  }
\end{figure}

PSB pair differencing is able to remove common-mode atmospheric 
fluctuations, 
and the relative gains are very stable (Figure~\ref{fig:diff}).
Even over a time scale of months, the relative gains are stable 
with $\sim$1\% rms measurement noise and exhibit no systematic variation 
with the optical loading.  Relative gains have also been derived from 
correlating timestream atmospheric fluctuations within PSB pairs, and 
although these have greater statistical uncertainties
the results are consistent with the elevation nods to within $\pm$3\%.

As an additional method to track gain variations, an infrared source 
supported by a foam paddle is swung into the beam to inject a signal of 
very stable amplitude, as described in \citet{Yoon2006}.  It produced a 
very repeatable (0.2\% rms) response between the beginning and end of 
the 1-hr scan sets and also showed that the individual gains are 
stable with 1\% rms across the full elevation range.  However, 
because of the $\sim$3~K optical loading introduced by the swing arm and 
unknown polarization of the infrared source, the relative gains from the 
flash calibrator have not been used for pair-differencing.

To quantify the level of leakage of the CMB temperature anisotropy into 
pair differences, the individual PSB pair-sum and pair-difference maps 
were cross-correlated.
This analysis performed on the yearly maps from the first 2 years showed no 
statistically significant evidence for relative gain errors in the data 
and placed an upper limit of $<$1.1\% rms on the angular scales of interest.
There is weak evidence that a small subset of PSB pairs, especially at 150 
GHz, have excess sum-difference correlations.
With the full 3-year data set, we anticipate placing tighter constraints 
on the gain uncertainties.
The power spectra of spurious $B$-mode polarization due to the 
best-estimate distributions of 1.1\%~rms relative gain errors are plotted 
in Figure~\ref{fig:bbsim}a.
Although these upper limits exceed the signal for $r=0.1$ on some scales, 
especially for 150~GHz, this systematic effect is still well below the 
statistical error in the first 2 years of data.  
If measured, this leakage can be mitigated by projecting out a CMB temperature 
anisotropy template from the polarization maps.

The measured individual gains are scaled for each 1-hr scan set such 
that the mean response of the detectors in each frequency band is constant.
Finally, the absolute gains for converting the detector voltage into CMB 
temperature units are derived by cross-correlating maps of CMB temperature 
anisotropy measured by \bicep\ and $WMAP$.  
As described in detail in \citet{Chiang2010},
the \bicep\ map (in volts) and ``\bicep-observed'' $WMAP$ maps (in K$_{\rm 
CMB}$) are identically smoothed and filtered, and we compute the absolute gain
$g_\ell$ for each of the frequency bands from cross-correlations in multipole 
space.
Within the multipole range of $\ell=$~56--265, $g_\ell$ is nearly flat, 
and the average is used as a single calibration factor for each frequency band.
We take the standard deviation of 2\% to be a conservative estimate of the 
absolute gain uncertainty in CMB temperature units.  
The results are consistent with those from the dielectric sheet calibrator 
(described in \S\ref{sec:polarization}), which can provide a real time 
absolute calibration with a 10\% uncertainty.

\subsubsection{Spectral response}

As described above, the relative gain calibration of a PSB pair is based 
on the relative response of the detectors to the change in atmospheric 
loading from a small nod in elevation.
Because the temperature derivative of the CMB and the atmospheric 
emission have different spectral shapes, the relative gain chosen to match
the response to elevation nods may not be optimal for the rejection
of CMB temperature fluctuations.

\begin{figure}[!htb]
 \centering
\includegraphics[trim=20 9 0 9,clip,width=\linewidth]{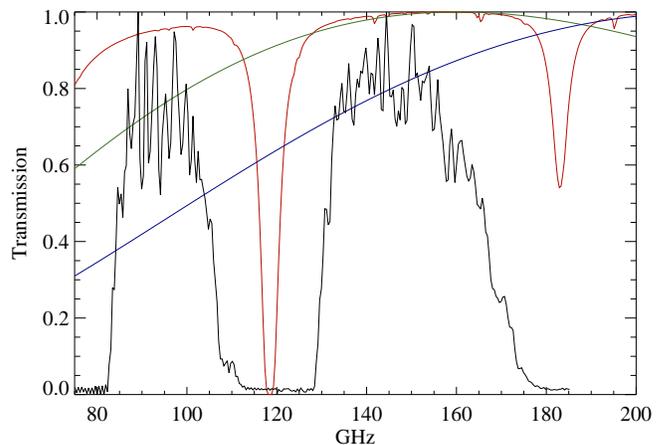}
 \caption[spectra]
 { \label{fig:spectra} Average measured spectral response for each 
of \bicep's frequency bands, normalized to unity.
Overplotted are the atmospheric transmission at the South Pole (red), the 
CMB spectrum (green) and its temperature derivative (blue).}
\end{figure}

The spectral response of each channel was measured using two separate
polarized Fourier Transform Spectrometers with a maximum resolution of
0.3~GHz, once in the lab and once in the field.  Within each frequency
band, the spectra were very similar from channel to channel (average
spectra shown in Figure~\ref{fig:spectra}), and the 
upper limit on the expected relative gain errors due to spectral 
mismatch was roughly 1\% rms over the array.
This current upper limit does not rule out spectral mismatch as a 
source of possible relative gain errors in the small subset of PSB pairs.

In addition to the main band, we verified that there is no significant
response at higher frequencies due to leaks in the low-pass filters. 
High-pass thick grill filters with cut-off frequencies of
165 and 255~GHz were used in front of the telescope aperture one
at a time and the response to a chopped thermal source was measured.
150~GHz channels showed no sign of leaks beyond 255 GHz down to the
noise floor at $-35$~dB, while 100~GHz channels exhibited leaks at
$\sim$ $-25$~dB level at $>$255~GHz.  The magnitude of this small
($\sim$0.3\%) leak was consistent between the PSBs in each pair, so
the effect on relative responsivities is expected to be negligible.

\subsection{Beam characterization} \label{sec:beams}

Mismatch in the beams of a PSB pair can result in a false polarization
signal from unpolarized temperature fluctuations.
Beam mismatches can also lead to the mixing between $E$-mode and 
$B$-mode polarization.
The difference between two nearly circular beams can be decomposed into 
three quantities corresponding to monopole, dipole, and quadrupole 
differentials: differential beam size $(\sigma_1-\sigma_2)/\sigma$,
differential pointing $({\bf r_1}-{\bf r_2})/\sigma$, and
differential ellipticity $(e_1-e_2)/2$,
where $\sigma_{1,2}$ are the Gaussian beam sizes of the first and second 
PSBs in a pair, $\sigma$ is the average beam size, and ${\bf r}_{1,2}$ are 
the centroid coordinates.
We define ellipticity as $e = (\sigma_a - \sigma_b) / (\sigma_a + 
\sigma_b)$, where $\sigma_{a,b}$ are widths of the major and minor axes, 
respectively.
Beam size and ellipticity differences are sensitive to the second spatial 
derivative of the temperature field, while pointing offset is also 
sensitive to the temperature gradient.  

For \bicep's focal plane layout and scan strategy, simulations show that 
differential beam size, pointing, and ellipticity of 3.6\%, 1.9\%, and 
1.5\% rms over the array, respectively, will result in spurious $B$-mode 
signal at the $r=0.1$ level.
The effect of differential pointing was simulated using the measured 
magnitude and direction of beam offsets with the expected amount of false 
$BB$ power scaling as the square of the magnitude.
Differential beam size was simulated by introducing a random distribution 
of beam size differences in PSB pairs, while keeping the average beam size 
the same.
Similarly, differential ellipticity was simulated by making the beam of 
every PSB elliptical by a small randomized amount such that the pairs have 
ellipticity differences of the given rms while keeping the beam sizes 
the same.  
The measurements described below indicate that the major axes tend 
to be more azimuthal with respect to the optical axis than radial, and the 
major axes of paired PSB beams tend to align within $\sim$15$\deg$ rms of 
each other.  Therefore, the major axes in the simulations were varied 
around the azimuthal direction by 10$\deg$ rms to roughly simulate the 
observed alignment trend within each pair.
Analytic calculations show, and these simulations verify, that the 
expected false $BB$ power scales as the square of differential beam size or 
ellipticity and as the fourth power of the beam size \citep{Shimon2008}.

\begin{figure}[!htb] \centering
\includegraphics[trim=85 20 100 40, clip, width=\linewidth]{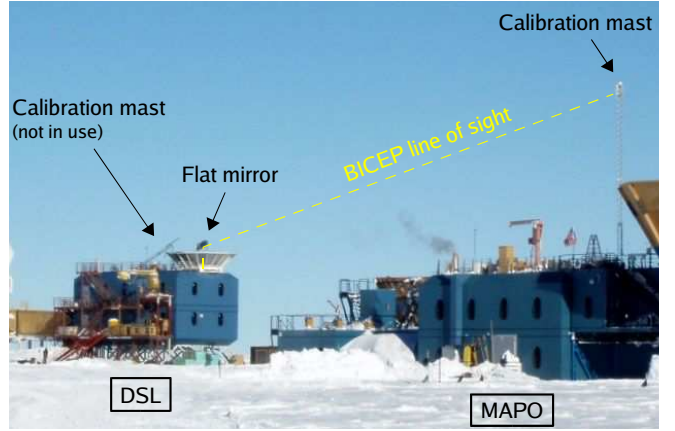}
\caption[beam_setup]
{ \label{fig:beam_setup} Beam mapping setup on site consisted of 
sources mounted on the top of fold-over masts.  When using the mast on the 
MAPO building (200~m from the Dark Sector Laboratory), a flat mirror is 
mounted to direct the beams over the ground screen.  }
\end{figure}

The beams were mapped by raster scanning a bright source at various 
boresight rotation angles. 
The far field of the telescope is about 50~m from the aperture, which 
permitted measurements in a high bay prior to telescope deployment as well 
as with the instrument installed at the South Pole.
In the high bay, a thermal blackbody source was used at a 40~m 
distance, consisting of a liquid nitrogen temperature load behind 
chopper blades covered with ambient temperature absorber.
At the South Pole, a temporary mast was installed on the rooftop 
outside of the fixed ground screen, allowing us to position a source at 
60$\deg$ elevation 10~m away.
For a truly far-field measurement, an additional mast was installed on 
the roof of the Martin A. Pomerantz Observatory (MAPO), at a distance of 
200~m, and a flat mirror was temporarily mounted on top of the 
telescope to direct the beams down to allow the observation of the distant 
mast as well as low elevation astronomical sources
(Figure~\ref{fig:beam_setup}).  The sources used included an ambient 
temperature chopper against the cold sky, a broadband noise source, the 
Moon, and Jupiter.
The broadband noise source is an amplified thermal source that is ideal 
for probing low-level effects.  

\begin{figure}[!ht] \centering
 \includegraphics[width=\linewidth]{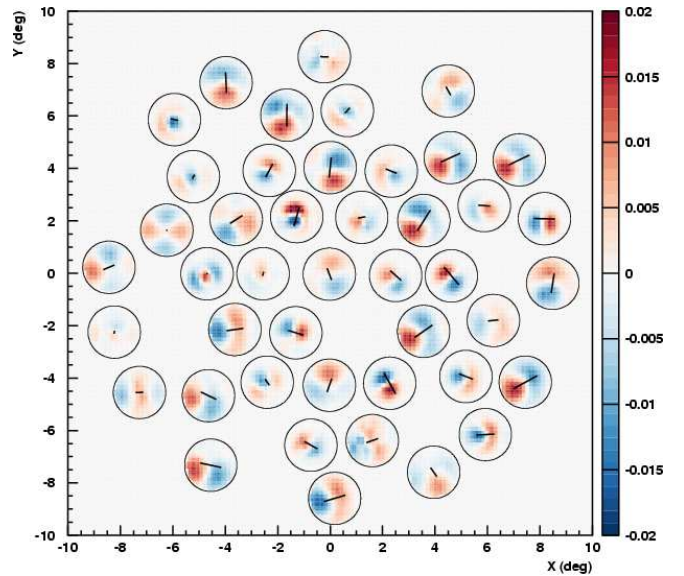}
 \caption[composite_diff] 
 { \label{fig:composite_diff} Beams for each PSB pair are
normalized and differenced to produce this composite differential beam
map (in the same orientation as Figure~\ref{fig:layout}).  The overplotted 
lines show the fitted pointing offsets magnified by a factor of 100.}
\end{figure} 

The measured beams are well fit with a Gaussian model, typically resulting 
in 1\% residuals in amplitude. The fitted centroids are repeatable to 
about 0.02$\deg$, 
although the accuracy of the absolute locations is currently limited by 
uncertainties in parallax and 
pointing corrections while using the flat mirror. 
The average measured FWHMs are 0.93$\deg$ and 0.60$\deg$ for 100 and 150 
GHz, respectively, about 5\% smaller than predicted from physical optics 
simulations.  The beam widths are measured to $\pm$0.5\% precision and 
vary by $\pm$3\% across the array.
The beams have small ellipticities of $e<$ 1\% at 100~GHz and $e<$ 1.5\% 
at 150~GHz.

The largest beam mismatch effect is a pointing offset that gives rise to 
dipole patterns in many of the differenced beams (Figure~\ref{fig:composite_diff}).  
The median differential pointing offset is 0.004$\deg$ at both 100 and 150 
GHz, and is on average 1.3\% of the beam size $\sigma$.  
The offsets were repeatable between observations of both the broadband 
noise source and the Moon to within the measurement uncertainty of 0.4\% 
of $\sigma$.

\begin{figure}[tb] \centering
 \includegraphics[width=\linewidth]{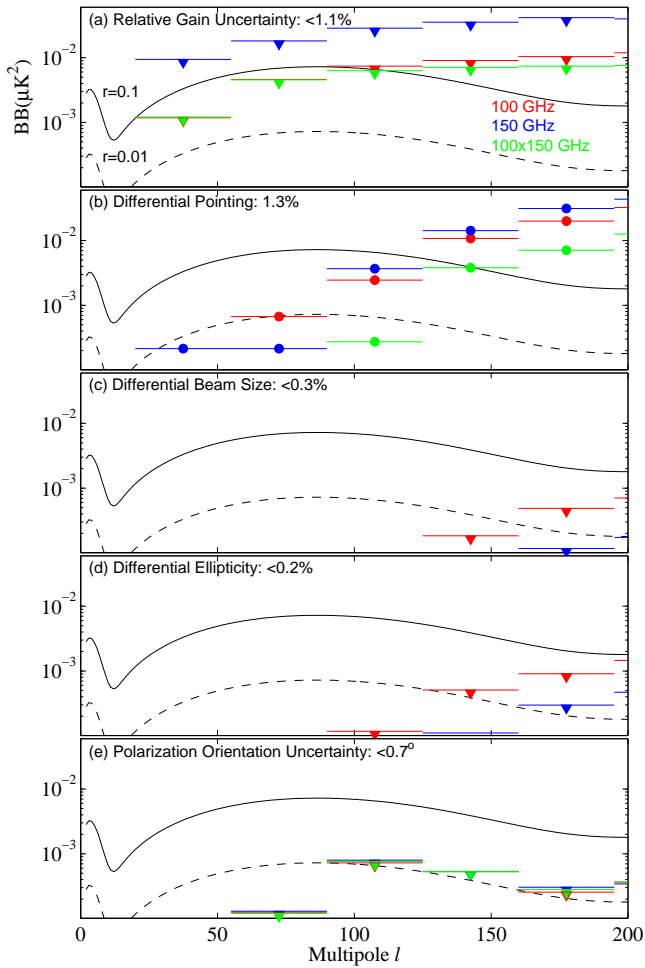}
 \caption[bbsim]
 { \label{fig:bbsim} Spurious $BB$ power from simulations of measured
potential systematic errors.  Except for differential pointing, all $BB$ 
estimates correspond to measured upper limits.
Effects of relative gain error and differential pointing can be corrected 
for in the analysis if necessary. 
All the potential systematic uncertainties are measured to be well 
below the 2-year constraint of $r<0.72$ \citep{Chiang2010}.}
\end{figure}

Simulated observations with the measured pointing offsets indicate a false 
$BB$ with an amplitude comparable to the $r=0.1$ spectrum 
(Figure~\ref{fig:bbsim}b), although well below the noise level of the 
initial 2-year data analysis.  
With the magnitude and direction of the pointing offsets measured 
precisely, the resulting leakage of CMB temperature gradients into 
polarization can be estimated and accounted for in future analysis.
Differential beam size and ellipticity are not measured with
significance; the measured upper limits of 0.3\% and 0.2\% rms, 
respectively, are negligibly small (Figure~\ref{fig:bbsim}c,d).

\subsection{Polarization orientations and efficiencies}
\label{sec:polarization}

To construct accurate polarization maps from PSB timestreams, we must know
the polarization orientation angle $\psi$ and cross-polarization response 
$\epsilon$ of each PSB.  
Accurate orientations must be used in map making to prevent 
rotation of $E$-mode polarization into false $B$-mode.
Cross-polarization response determines the polarization efficiency 
$(1-\epsilon)/(1+\epsilon)$, which affects the amplitude scaling of the 
power spectrum.
We developed experimental techniques to measure these quantities 
by injecting polarized radiation into the telescope aperture at many 
different angles with respect to the detectors.  The phase and amplitude 
of each PSB's response determine $\psi$ and $\epsilon$, respectively.
This section discusses the calibration benchmarks for these quantities and 
describes three measurement techniques and their results.
The absolute PSB orientations were measured to within $\pm0.7\deg$ and 
relative orientation to within $\pm0.1\deg$, and $\epsilon$ was measured 
to within $\pm$0.01.

Angles of the PSBs can vary from their design orientations due to the 
limited mechanical tolerances with which they are mounted.
The deviation from perfect orthogonality of a pair simply reduces its 
efficiency for polarization; however, an error in the overall orientation 
of the pair can lead to rotation of $E$-modes into $B$-modes.  
With the expected fractional leakage being sin(2$\Delta\psi$), 
the $\sim$1 $\mu$K $E$-modes at $\ell=100$ can rotate into false $B$-modes 
at the $r=0.1$ level of 0.08 $\mu$K if the orientation measurement is off 
by 2.3$\deg$.  This benchmark and the expected scaling were verified by 
simulations of systematic orientation offset of all the PSBs.
The calibration procedure was designed to determine the polarization 
orientations to within a degree.

Another factor, though less important, is that the PSBs are not perfectly 
insensitive to polarization components orthogonal to their orientations, 
effectively reducing the polarization efficiency to $(1-\epsilon)/(1+\epsilon)$.
To achieve 10\% accuracy in the amplitudes of the polarization power 
spectra, which are proportional to $(1-\epsilon)^2/(1+\epsilon)^2$, our 
goal was to measure cross-polarization responses $\epsilon$ to 
better than $\pm$0.026.

\begin{figure}[!htb]
\centering\includegraphics[width=\linewidth]{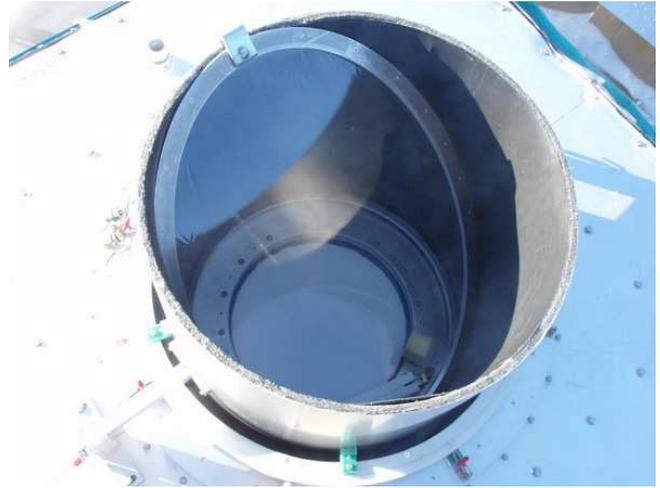}
\caption[calibrator]
{\label{fig:calibrator} Dielectric sheet calibrator for measuring PSB 
orientations consists of a beam-filling polypropylene sheet and an ambient 
load made of a highly emissive black lining, subjecting the beams to 
partially polarized radiation.
The device is mounted on the azimuth stage, which can rotate about the 
telescope's boresight when pointed at zenith.}
\end{figure} 

The polarization orientations were measured using a rotatable dielectric 
sheet (Figure~\ref{fig:calibrator}), modeled after the one used by 
POLAR \citep{ODell2002}.  
A small partially polarized signal of known magnitude is created by using 
an 18-$\mu$m polypropylene sheet in front of the telescope aperture 
oriented at 45$\deg$ to the optical axis. 
The sheet acts as a beam splitter transmitting most of the sky radiation 
but reflecting a small polarized fraction of the radiation from an ambient load 
perpendicular to the beam.  
The polarized signal is small compared to the unpolarized sky background 
so that it can provide an absolute responsivity calibration in optical 
loading conditions appropriate for normal observations.
The ambient load is made of a microwave absorber 
lining inside an aluminum cylinder surrounding the beam splitter.  
The absorber is covered with a 1/8" thick sheet of closed cell expanded 
polyethylene foam
exactly as in the forebaffle (described in \S\ref{sec:sidelobe}), the 
combination of which has $\sim$95\% emissivity at 100 GHz.  

\begin{figure}[!htb]
\centering\includegraphics[trim=10 5 5 10,clip,width=\linewidth]{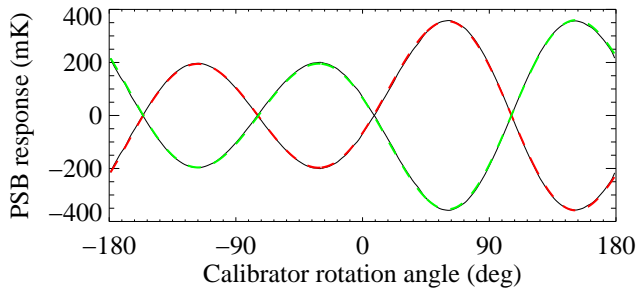}
\caption[polcal]
{\label{fig:polcal} Response of each PSB in a pair as a function of 
the dielectric sheet calibrator orientation (black), plotted over by the 
fits (in red/green dashes) with polarization orientations and 
responsivities as free parameters.}
\end{figure}

We use this polarization calibrator by putting it in the place of the forebaffle 
and fixing it to the azimuth mount.  With the telescope pointed at zenith, 
rotating the device with respect to the cryostat modulates the 
polarization signal for each detector while keeping the beams stationary 
with respect to the sky.  
The off-axis beams see complicated, but calculable, deviations from the 
nominal sinusoidal modulation (Figure~\ref{fig:polcal}).
This setup produces a partial polarization of amplitude proportional to 
$(T_{\rm amb}-T_{\rm sky})$, the temperature difference between the 
ambient load and the sky loading.
With an 18-$\mu$m polypropylene film and a typical temperatures of 
$T_{\rm amb}$ = 220 K and $T_{\rm sky}$ = 10 K, the signal amplitude is 
$\sim$100 mK at 100 GHz and $\sim$250 mK at 150 GHz, small enough to 
ensure that the bolometer response remains linear.

The measurements were performed several times throughout each observing 
year and produced repeatable results for the individual PSB orientations 
with 0.1$\deg$ rms.  
The relative orientation uncertainty of $\Delta(\psi_1-\psi_2)=0.1\deg$
results in negligible absolution calibration error.
The PSB pairs were found to be orthogonal to within 
0.1$\deg$, and together were within 1$\deg$ of the design 
orientations shown in Figure~\ref{fig:layout}.  
PSB orientation measurements performed before and after the focal plane 
servicing (2006 November) show a possible discrepancy, corresponding to an 
average of 1.0$\deg$ global rotation.  We therefore conservatively assign 
a $<$0.7$\deg$ rms uncertainty in the absolute orientation for each year.
This absolute orientation angle accuracy is sufficient for measuring 
$r\ll0.1$ (Figure~\ref{fig:bbsim}e).

\begin{figure}[!htb]
\centering\includegraphics[width=\linewidth]{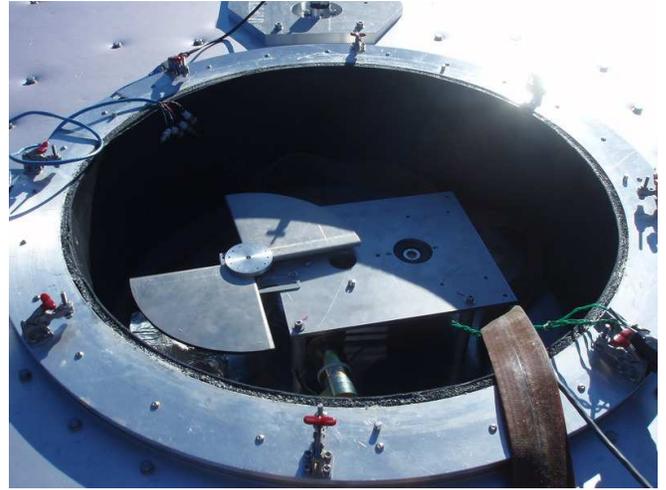}
\caption[aperture]
{\label{fig:aperture} Device above the cryostat window for measuring 
cross-polarization responses and PSB orientations.  The window is covered 
with a metal plate with a 2-cm Eccosorb aperture, and a 10-cm diameter 
wire grid is on a rotation stage under the circular aperture of the 
rectangular plate.  The chopper modulates the load between the ambient 
temperature and the cold sky.  }
\end{figure}

The cross-polarization responses $\epsilon$ were measured using two 
methods that also independently confirmed the absolute orientation 
measurement.  
One method used a rotatable wire grid in front of the telescope window 
with a chopper modulating the polarized load through a small 
aperture between the ambient absorber and the cold sky 
(Figure~\ref{fig:aperture}).  Fitting a sinusoid to the individual PSB 
response as a function of the wire grid angle gives the polarization 
efficiency and orientation.
The measurements for all three years gave cross-polarization response 
values with a distribution $\epsilon=0.045 \pm0.02$. 
One 150 GHz bolometer was omitted from analysis for having an 
$\epsilon>0.12$ and was replaced at the end of the first year.

\begin{figure}[!htb]
\centering\includegraphics[width=\linewidth]{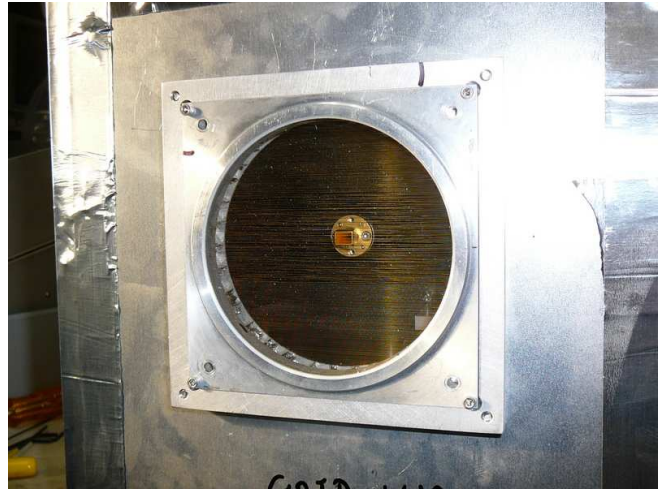}
\caption[bns]
{\label{fig:bns} Another calibration source, used on top of a mast, for 
measuring cross-polarization responses (and PSB orientations).  The 
broadband noise source at 100 or 150 GHz outputs power through the 
rectangular feedhorn oriented for either 
vertical or horizontal polarization and through a precisely aligned wire 
grid to minimize cross-polarized signal. }
\end{figure}

The other method used a modulated broadband noise source with a 
rectangular horn behind a wire grid, mounted on the mast 200~m away 
(Figure~\ref{fig:bns}).  This source was raster scanned by each of our 
beams with 18 different detector orientations with respect to the wire 
grid, fitting a 2-dimensional Gaussian to each raster.  The measured 
cross-polarization responses were slightly lower with a median of 
$\epsilon=0.038$.
Based on the scatter against the results from the first method, we assign 
an uncertainty of $\Delta\epsilon=0.01$, which translates to 4\% 
uncertainty in the polarization power spectrum amplitudes.

\subsection{Telescope and detector pointing} \label{sec:pointing}

Pointing errors greater than 1\% of the beam size $\sigma$ could 
contaminate the $B$-mode spectrum at the $r\sim10^{-4}$ level 
\citep{Hu2003}.  
Since the amount of spurious $BB$ power scales as the square of the pointing 
error, the $r=0.1$ benchmark would correspond to 0.3$\sigma$, or 5$^{\prime}$ for 
\bicep.
This benchmark was verified to be conservative by simulating a
5$^{\prime}$ rms shift in boresight pointing every 10 elevation steps in 
one of our simulation pipelines and finding negligible spurious signal 
compared to the $r=0.1$ $BB$ signal.
An optical star pointing camera is used to measure the telescope boresight 
pointing model with uncertainties more than an order of
magnitude smaller than required to achieve our benchmark. 

An accurate derivation of the sky coordinates from the telescope encoder 
readings requires a precise knowledge of the 
state of the mount, including axis tilts, encoder offsets, and flexure.
Encoder data for the three mount axes are recorded synchronously
with the bolometer timestreams and corrections to the raw pointing data 
are applied during map making using a pointing model.
The pointing model is established using a compact optical 
star-pointing camera with a 2$^{\prime\prime}$ resolution 
mounted beside the main window on top of the cryostat and co-aligned with 
the boresight rotation axis.
There are 10 dynamic parameters: AZ axis tilt magnitude and 
direction, EL axis tilt, the 3 encoder zeros, amount of telescope flexure 
$\propto$ cos($EL$) and $\propto$ sin($EL$), and magnitude and direction of the 
collimation error of the pointing camera itself.  A complete 
characterization of these parameters requires the observation of at least 20 stars.
To establish a pointing model during the Antarctic summer, the 
pointing camera was designed to be sensitive enough to detect magnitude +3 
stars in daylight.
For maximum contrast against the blue sky, we used a sensor with enhanced 
near-infrared sensitivity\footnote{Astrovid StellaCam EX with EXview HAD 
CCD, http://www.astrovid.com/prod\_details.php?pid=7}
and an infrared filter cutting off below 720~nm.
We used a 100-mm diameter lens that was color-corrected and 
anti-reflection coated for 720--950~nm.
Its 901-mm focal length results in a small 0.5$\deg$ field of view to keep 
the sky background low. 
Careful adjustments of the CCD camera and the mirrors reduced the optical 
camera's collimation error to 2.4$^\prime$.  During our first Antarctic 
summer season, we successfully captured 26 stars down to magnitude +2.9 
in an elevation range of 55$\deg$--90$\deg$.

Optical pointing calibrations were performed every two days during the 
refrigerator cycles, weather permitting, as well as before and after each 
mount re-leveling.
In each run, the telescope is pointed at 24 stars at boresight 
rotation angles of $-45\deg$, $45\deg$, and $135\deg$, and the 
azimuth and elevation offsets required to center each star are recorded.  
The pointing data are fit to an 8-parameter model
(it has not been necessary to fit for telescope flexure) 
with typical residuals of $0.2^\prime$ rms. 
The pointing model has been checked by cross-correlating the CMB 
temperature anisotropy patterns between the pointing-corrected daily maps 
and the cumulative map; no systematic offsets or drifts are detected.

In parallel with the optical pointing calibration, the tilt of the 
telescope mount is monitored every two days using two orthogonal tilt 
meters mounted on the azimuth stage.  We observed seasonal tilt changes of 
up to 0.5$^\prime$ per month, possibly due to the building settling on 
the snow, and typically re-leveled the mount before the tilt exceeded 
1$^\prime$.

Finally, to co-add maps made with different PSB pairs, the actual 
locations of all the beams relative to the boresight must be determined.  
This was accomplished by first making a full season co-added map of the 
CMB using the design locations and then cross-correlating the temperature 
anisotropy pattern with single detector maps for each of the four boresight 
rotation angles to adjust the individual beam coordinates.  These adjusted 
coordinates with respect to the boresight were then used to iterate this 
process until every individual pair map was consistent with the full co-added 
map. 
This derivation of the absolute beam locations resulted in
an uncertainty of $2^{\prime}$ rms, based on the agreement between the 
first and second years.  For \bicep, using the CMB temperature 
fluctuations proved to be more effective than attempting a similar procedure 
with Eta Carinae, the brightest compact source accessible.

\subsection{Sidelobe rejection} \label{sec:sidelobe}
Sidelobes of the telescope beams can pick up emission from the bright 
Galactic plane and structures on the ground, possibly resulting in 
contamination of the polarization maps.
The ground shields were designed to reject the ground radiation to a level 
where the contamination is below our target $B$-mode polarization sensitivity.  
This section describes the ground shield design, the sidelobe 
measurement, and the possible polarization contamination due to the 
sidelobes.
\bicep's sidelobes are sufficiently low to enable the measurement of 
$B$-mode polarization to the level of r=0.01.

\begin{figure}[!htb]
\centering\includegraphics[width=\linewidth]{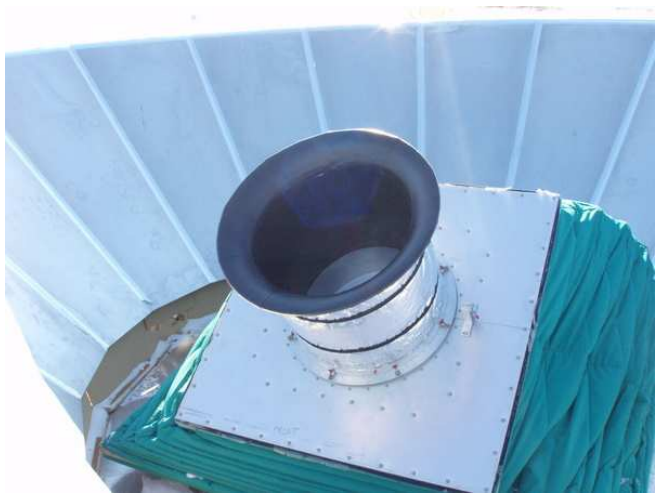}
\caption[shields]
{\label{fig:shields} Absorptive forebaffle and the reflective ground 
screen.}
\end{figure}

\bicep\ uses two levels of shielding against ground radiation: an 
absorptive forebaffle fixed to the cryostat and a large stationary 
reflective shield surrounding the telescope structure 
(Figure~\ref{fig:shields}), modeled after the POLAR experiment 
\citep{Keating2003b}.  
We designed the geometry so that any radiation from the ground must be 
diffracted at least twice before entering the window in any telescope 
orientation. 

The forebaffle is an aluminum cylinder lined with a microwave absorber to 
minimize reflected radiation into the telescope.  
It is wide enough to clear 
the edge pixel beams, beyond which the Zotefoam\footnote{Propozote PPA30, 
http://zotefoams.com/pages/en/datasheets/PPA30.htm}
window \citep{Runyan2003} is expected to scatter $<$1\% of the total 
power.  
At \bicep's lowest nominal CMB observing elevation of 55$\deg$,
the forebaffle is long enough to prevent radiation from sources, 
particularly the Moon, at elevations up to 27$\deg$ from entering the 
window directly.  
The forebaffle aperture lip is rounded with a 13~cm radius to reduce 
diffraction of the diffuse beam sidelobes.

After testing many materials for the microwave absorber, we chose a 10-mm
thick open-cell polyurethane foam sheet (Eccosorb 
HR\footnote{http://www.eccosorb.com/america/english/product/40/eccosorb-hr}),
which had the lowest measured reflectivity of $<$3\% at 100 and 150~GHz 
when placed over a metal surface (150~GHz results by W. Lu \& J. Ruhl 
2004, private communication).
To prevent snow from accumulating in the porous Eccosorb foam, it is 
covered with a 1.6-mm thick smooth polyethylene foam 
(Volara\footnote{Volara; 
Sekisui Voltek, http://www.sekisuivoltek.com/products/volara.php}), 
which is attached with a silicone sealant.
The combined Eccosorb HR~/ Volara stack was measured to reflect $\sim$5\%
of $100\,$GHz radiation incident at 45$\deg$.  
The additional loading on the bolometers due to emission from the 
forebaffle was measured to be $\sim$1 K$_{\rm RJ}$.  
Since the absorptive baffle is fixed with respect to the detectors, its 
thermal emission is expected to be stable against the modulated sky 
signal.  

The 2-m tall outer screen prevents the forebaffle lip from seeing
the warm ground.  The sloped aluminum surface instead reflects any 
diffuse sidelobes to the relatively homogeneous cold sky.  The
8-m top diameter is wide enough so that the diffracted ground
radiation will never directly hit the window even when the telescope is at 
its 50$\deg$ elevation lower limit. 
The edge of the outer screen is also rounded with
a 10~cm radius to reduce diffraction.

The sidelobe response of the telescope, including the forebaffle, was 
measured using a chopped broadband noise source on the mast $10\,$m from 
the telescope aperture. 
The telescope was stepped in elevation up to 60$\deg$ away from the 
source in 0.5$\deg$ increments, making one revolution about the boresight 
and back at each step to measure a radial average of the beam.  
This measurement was performed with several source attenuations down to 
below $-50$~dB to probe the far sidelobes with sufficient signal-to-noise 
ratio while also measuring the main beam without saturating the detector.

\begin{figure}[!hbt]
 
\centering\includegraphics[trim=0 10 0 10,clip,width=\linewidth]{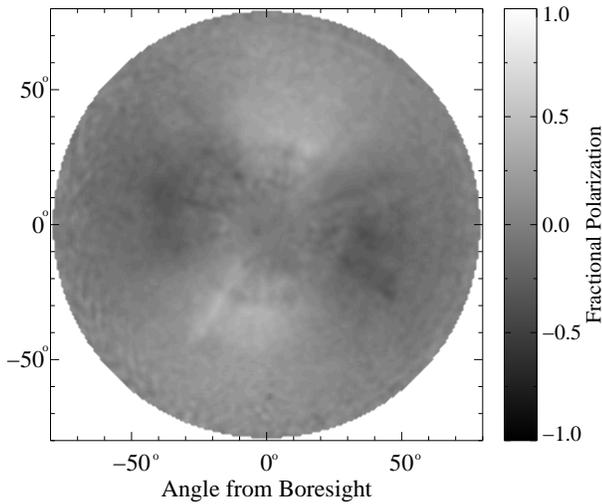}
 \caption[sidelobe_pol]
 { \label{fig:sidelobe_pol} 
Map of fractional polarization for a 100-GHz feed ("C1") up to 80$\deg$ 
from the beam center, showing the gain-adjusted pair difference divided by 
the pair sum.  Beyond $\sim$15$\deg$ from the boresight where the 
forebaffle cuts off the beam, the sidelobes are generally up to 50\% 
polarized, but with smooth quadrupole pattern aligned with the 
polarization sensitivity. } 
\end{figure}

\begin{figure}[!htb]
 \centering\includegraphics[angle=90,trim=0 0 0 0,clip,width=\linewidth]{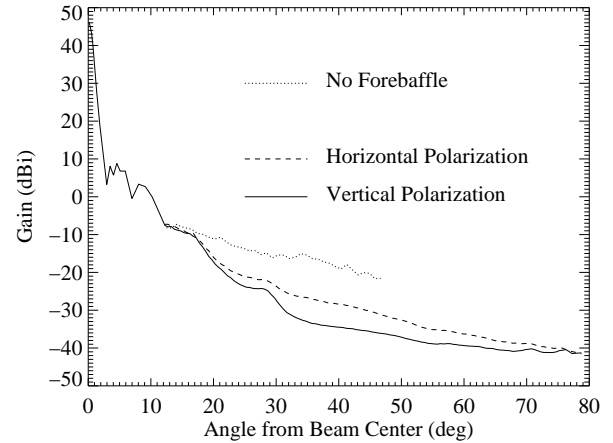}
 \caption[sidelobe] 
 { \label{fig:sidelobe} Azimuthally-averaged sidelobe response for the 
100-GHz central feed.  
A forebaffle with an absorptive lining cuts in at 15.5$\deg$ and provides 
up to an additional $\sim$15~dB attenuation.  When the telescope is at its 
lowest elevation of 50$\deg$, the lip of the outer ground screen is 
$\sim$30$\deg$ from the beam center of the central feed.  }
\end{figure}

Sidelobe response maps constructed from the gain-adjusted 
pair differences show that the sidelobes are up to 50\% polarized 
(Figure~\ref{fig:sidelobe_pol}).
The individual sidelobe maps can be averaged over boresight rotation angle
to obtain a radial profile (Figure~\ref{fig:sidelobe}).  
Waves polarized parallel to the absorbing forebaffle lip surface are more 
strongly diffracted than those polarized perpendicular to the lip.
This difference appears to be mainly responsible for the polarized 
response in the sidelobes.

\begin{figure}[!htb]
\centering\includegraphics[angle=90,width=\linewidth]{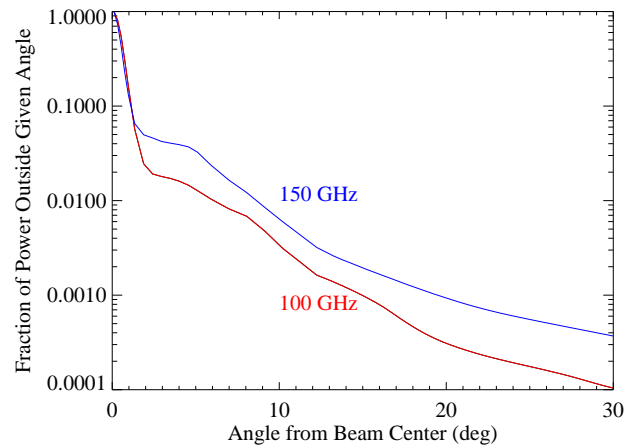}
 \caption[integratedbeam] 
 { \label{fig:integratedbeam} 
The sum of the beam profiles for the horizontal and vertical polarizations 
in Figure~\ref{fig:sidelobe} is integrated over the solid angle from 
80$\deg$ to the given angle, and the fraction with respect to the total 
power is plotted.
Less than 0.1\% of the power in the beam is left beyond 15$\deg$ and 
20$\deg$ of the beam center at 100 and 150~GHz, respectively. }
\end{figure}

To quantify what fraction of the total power in the beam remains outside 
of a given angle from the beam center, the net beam profile for horizontal 
and vertical polarizations was integrated 
(Figure~\ref{fig:integratedbeam}).
Less than 0.1\% of the power in the beam is left beyond 15$\deg$ and 
20$\deg$ of the beam center at 100 and 150 GHz, respectively.

Polarized sidelobes can result in spurious signals by coupling to emission 
from the Galaxy and the outer ground screen itself.
To evaluate the far sidelobe rejection performance, the measured level of 
polarized response was convolved with a model map of a potential 
contaminant, and an angular power spectrum was computed in the \bicep\ 
field to compare with the $r=0.1$ $BB$ spectrum.
For the Galactic model, we used the dust emission predicted by 
\citet{FDS1999}.
The resulting $B$-mode contamination was found to be at least 400 times 
below the $r=0.1$ level, meaning the measured sidelobes are at least 13 
dB below the benchmark.
The same exercise was repeated for a conservative model of snow 
accumulation on the ground screen panels.
The contamination was even smaller, with the achieved rejection level
at least 23~dB better than the benchmark.

We have also probed potential ground contamination in our data through a 
jackknife test comparing maps made in different azimuth ranges, and have 
seen no evidence of ground signal.
As mentioned in \S\ref{sec:strategy}, the scan range is fixed with respect 
to ground during each 1-hr scan set so that subtracting a 
scan-synchronous template each hour removes any ground-fixed signal.

\subsection{Thermal stability} \label{sec:thermal}

The thermal and optical responsivities of PSBs in each pair are not 
perfectly matched, so the temperatures of the detector focal plane and 
the emissive optics must be sufficiently stable to prevent the introduction 
of scan-synchronous thermal signals.
We have measured the thermal responsivity of every bolometer and compared 
the mismatches with the focal plane temperature stability, which we 
control with a feedback loop.
The thermal stability of both the focal plane and the optics is found to 
be adequate compared to the $r=0.1$ benchmark.

The bolometers' responsivities to the bath temperature are measured by 
correlating the detector timestreams with the 10~mK drop when the 
temperature control heater is turned off at the end of each refrigerator 
cycle.
The median thermal responsivity, after converting voltages into CMB 
temperature units, is 0.8~$\mu$K$_{\rm CMB}$/nK$_{\rm FP}$, and the 
median mismatch within PSB pairs is 0.08~$\mu$K$_{\rm CMB}$/nK$_{\rm FP}$.  
Because the pair differential responsivities
are distributed randomly in the array, 
the effects of the mismatch will average out when the maps are co-added.  
The averaged mismatch over the array, considering both the magnitudes and 
signs of the thermal response, 
is 0.025 and 0.001 $\mu$K$_{\rm CMB}$/nK$_{\rm FP}$ at 100 and 150 GHz, 
respectively.  
To meet the $r=0.1$ target of 0.08~$\mu$K$_{\rm CMB}$ at $\ell$$\sim$100, 
thermal instabilities in the focal plane must then be controlled to better than 
3~nK rms.  

To mitigate the thermal fluctuation effects, the focal plane temperature 
is stabilized at 250~mK using a 100~k$\Omega$ resistor as a control heater 
(nominally depositing $\sim$0.1~$\mu$W) in a PID feedback loop with a 
sensitive NTD germanium thermistor. 
The PID parameters are set such that no active regulation takes place 
within the observational signal band of 0.1--1 Hz; only long time-scale 
drifts are controlled so that the PSB relative gains remain unaffected. 
The focal plane is equipped with 6 pairs of monitor thermistors spaced 
evenly around its perimeter.
During the first year, the thermal control scheme used a thermistor 
closest to the thermal strap connected to the refrigerator.
To improve the recovery time from major thermal disturbances and other 
transient events, additional control thermistors were installed near 
the control heater on the thermal strap at the end of the first year. 
The rigidity of the thermal straps were improved by using 
Vespel\footnote{http://www2.dupont.com/Vespel/en\_US} 
supports to reduce susceptibility to vibrationally induced heating. 
Along with an increased response speed of the control loop, the temperature
stability measured by the focal plane thermistors was improved 
from the first year to the second.

The temperature stability of the focal plane was found to vary with 
azimuth scan speed as well as the cryostat orientation about its axis.  
The stability was investigated under a variety of telescope operating 
conditions, including scan speeds in a range of $1.0\deg$/s -- 
$4.0\deg$/s, and 16 evenly spaced boresight rotation angles. 
Based on the minimum variance of the scan-synchronous thermistor signals, 
we selected the 2.8$\deg$/s nominal scan speed and four cryostat 
orientations about the boresight: $\{-45\deg, 0\deg, 135\deg, 180\deg\}$.
The measured level of thermal fluctuations at the \bicep\ focal plane 
was 1~nK rms in the frequency range corresponding to $\ell=100$.

Finally, since emission from \bicep's optics is expected to be largely 
unpolarized, the main concern with optics temperature drifts is in 
mis-calibration of PSB pair optical relative gains, which have an upper 
limit of 1.1\% rms, as described in \S\ref{sec:elnod}.
To limit the pair-difference response to less than 0.08~$\mu$K$_{\rm CMB}$, 
the scan-synchronous optical loading fluctuations must be under 
$\sim$8~$\mu$K$_{\rm CMB}$, requiring the optics temperature to be stable 
to at least 4~$\mu$K$_{\rm RJ}$ rms (for the 150~GHz band).
Scan-synchronous fluctuations averaged over all the individual bolometers 
for a two month period show 0.7~$\mu$K$_{\rm RJ}$ rms variation 
in the frequency range 0.1--1 Hz.

\section{Noise properties and modeling} \label{sec:noise}

Precise 
characterization of noise in the \bicep\ data is crucial for accurately 
extracting the underlying CMB polarization angular power spectrum.
The detector timestreams are dominated by noise 
which must be modeled and simulated to subtract the noise bias
from the resulting power spectra. 
Precise subtraction is critical because any misestimation results in a 
systematic error in the power spectrum amplitude.
Simulating the effect of a noise misestimate on the final $BB$ spectrum from 
the 2-year \bicep\ data, we find that a $\pm 3\%$ overall error in the 
noise power estimate would result in a bias of $r=\pm0.1$.
This translates to a benchmark of 1.5\% accuracy in CMB temperature units 
for the pair difference noise simulation.
Also, the simulation of noise, along with that of signal, determines 
the error bars of the CMB power spectra and the constraints
on the $B$-mode polarization amplitude.
To remove noise bias from the raw power spectra, we simulate
 signal-free, noise-only timestreams
 and process them with the same pipeline as the actual data
 to compute the noise power spectra.
This section describes characterization of the noise properties, 
simulation of noise-only timestreams, and the resulting estimates of the 
noise bias in the final angular power spectrum.
The noise level is consistent with expectations and 
no significant cross-correlations in noise are observed among pair 
differences, and the simulated noise has been checked for consistency with 
the actual noise level in the data.

\subsection{Properties of noise: spectra and covariance}

To simulate noise-only timestreams with the same statistical properties as the 
actual data, the noise properties must first be modeled.
Because \bicep's raw timestreams are dominated by noise, we model the noise by 
simply computing the auto and cross spectral power distributions for 
detector sums and differences.
The signal-to-noise ratio in the timestreams is $\leq$0.2\% for the PSB 
pair differences and 1--10\% for the pair sums in the 0.1--1~Hz frequency 
range corresponding to $\ell=$~30--300.
The significant CMB signal in the pair-sum timestreams is expected to 
introduce some error, but we permit this in the present analysis because 
the uncertainty in noise bias is expected to be much smaller than the cosmic variance
in the temperature power spectrum. 
The present noise model accounts for correlations among all the detectors 
in the focal plane, but does not attempt to include correlations between 
half-scans or between orthogonal Fourier modes within each half-scan.  

PSB pair sums and differences are modeled and simulated after the removal 
of a third-order polynomial from the 20~s of each half-scan 
that are used in the CMB map making.
For each pair sum or difference consisting of 200 points at a 10~Hz 
sampling rate, we take the Fourier transform and multiply by its complex 
conjugate for an auto power spectrum in frequency space.  
We multiply the Fourier transforms of different pair sums or differences 
during each half-scan,
${\tilde d_A} {\tilde d_B}^* (2/\Delta\nu)$,
to obtain complex cross spectra in units of V$^2$/Hz,
where $\Delta\nu$ = 0.05~Hz is the frequency resolution.

\begin{figure}[!ht]
 \centering\includegraphics[width=\linewidth]{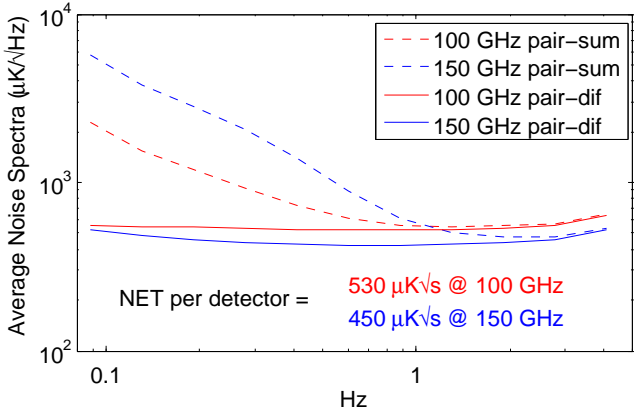}
 \caption[noisespectra]
 { \label{fig:noisespectra} Average noise power spectra of all PSB pairs 
used in the analysis, averaged over all observing blocks during the entire 
first 2 years.
The pair-sum spectra show 1/$f$ atmospheric noise, which is rejected by 
pair differencing.
NETs per detector are derived from the pair-difference average between
0.1 and 1~Hz.
Accounting for polarization efficiencies,
these correspond to an average instantaneous (i.e., single Stokes parameter)
``NEQ per feed'' of 410 and 340 $\mu$K$\sqrt{s}$ for 100 and 
150 GHz, respectively.  }
\end{figure}

\begin{figure}[!ht]
 \centering\includegraphics[width=\linewidth]{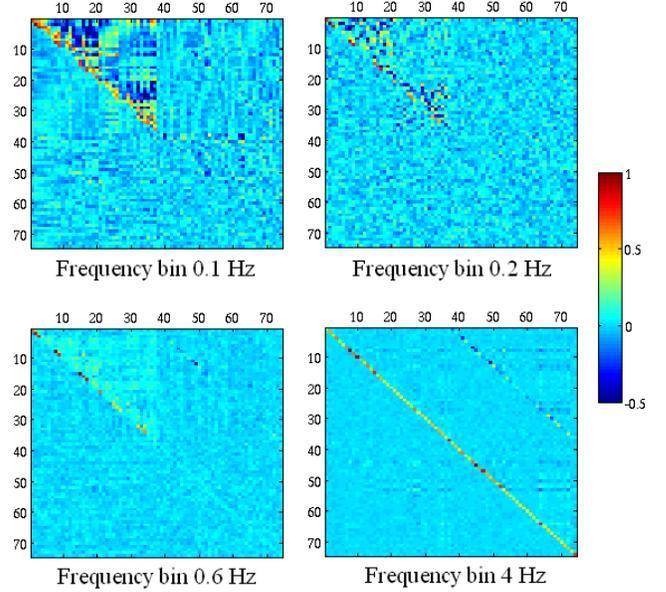}
 \caption[noisematrix]
 { \label{fig:noisematrix} Example correlation matrices from the \bicep\ 
noise model.  Each panel shows the fractional cross-correlations among PSB 
sums (indexes 1--37) and differences (indexes 38--74) of 37 PSB pairs 
used in 2007, with the real and imaginary components plotted in the 
upper-right and lower-left halves, respectively.  
The auto-spectra values along the diagonal are normalized simply by the 
maximum of the 74 values.
The atmosphere-induced correlations are visible as off-diagonal 
structure among the pair sums.
The imaginary component lacks substantial power, except among the pair 
sums at the lowest frequencies where the correlations due to atmosphere 
can be phase shifted depending on the relative beam locations.
We define the noise model in 12 frequency bands spanning 0.05--5 Hz, 
and four of those bands are shown.}
\end{figure}

For each 1-hr constant-elevation scan set, we average the complex 
spectra of 100 half-scans. 
By comparing the average spectra over the first and the second halves (50 
half-scans each) of the scan set, we checked that the noise properties are 
approximately stationary during a 1-hr period compared to the 
uncertainties in the averaged spectra.
Similarly, average spectra for right-going and left-going scans were 
compared during a scan set to verify that there is no significant 
difference.
To compute the noise model, we then bin the spectra into 12 
logarithmically spaced frequency bands spanning 0.05--5~Hz.
The average auto spectra over all PSB pairs in each frequency band are 
plotted in Figure~\ref{fig:noisespectra}, with the detector voltages 
converted to CMB temperature differences.
The NETs are consistent with expectations, 
and are comparable to the best achieved in other ground-based experiments 
\citep{Runyan2003, Hinderks2009}.
Before differencing, the timestreams show significant atmospheric noise 
below 1 Hz.
This is effectively rejected by pair differencing, although there is a hint of 
excess low frequency noise, especially at 150~GHz.
All auto and cross spectra for pair sums and pair differences are 
combined to form a complex noise covariance matrix ${\bf { \tilde N}}(f)$ 
at each of the 12 frequency bands, some of which are shown in 
Figure~\ref{fig:noisematrix}.

For the purpose of generating the noise model, the pair sum and difference 
timestreams are gap-filled for cosmic ray hits and electronic glitches.  
This procedure is not performed in making the CMB maps; any detector 
half-scan with a glitch is simply excluded. 
Because the noise covariance matrices are constructed by averaging auto 
and cross spectra over multiple half-scans, excluding a half-scan for a 
single detector pair sometimes causes the matrices to become 
non-positive-definite, which prevents the Cholesky decomposition necessary 
for the timestream simulation process described in the following section.  
Excluding a half-scan for all detectors if any of them contains a glitch 
results in data loss of up to 70\% for the noise model calculation.  
We therefore fill gaps when possible, and reject a half-scan for all 
detectors if more than four PSB pairs display a simultaneous glitch.

\subsection{Simulation of noise-only timestreams}

For each 1-hr scan set, the measured noise covariance matrix 
is used in the following steps to generate simulated noise 
timestreams that reflect the modeled noise correlations.

\begin{figure*}[!hbt]
\resizebox{\textwidth}{!}{\includegraphics{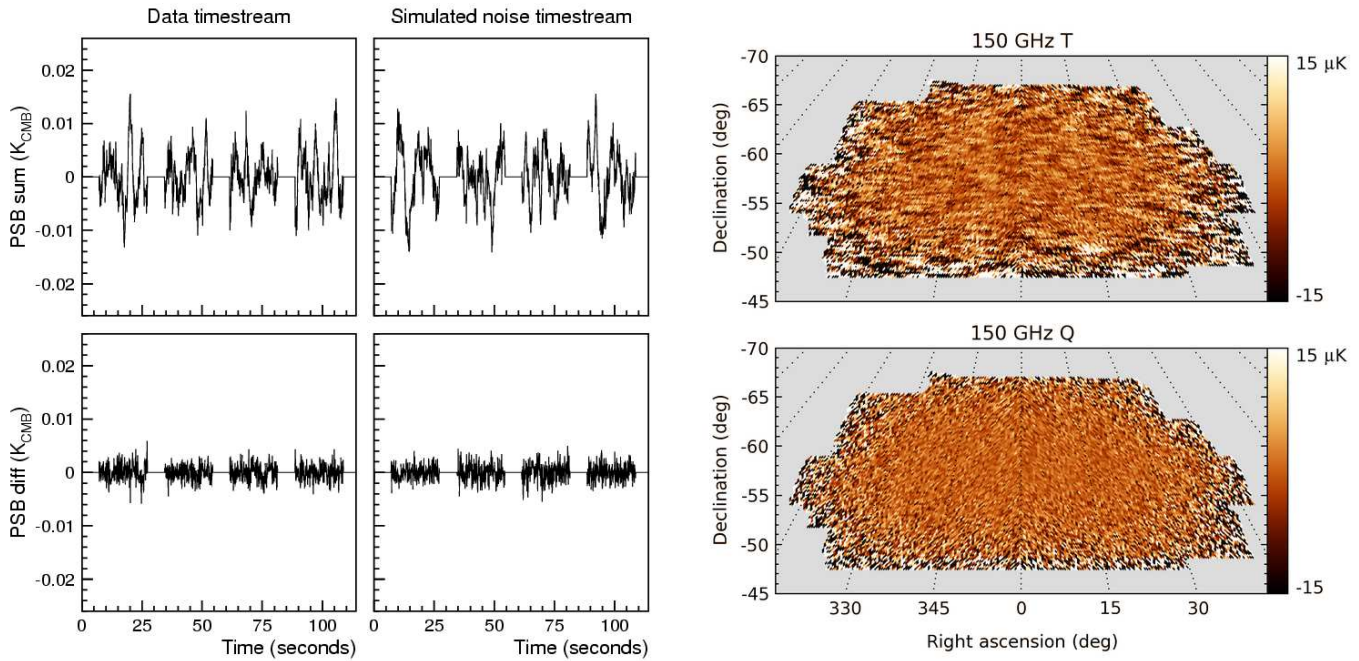}} 
\caption{\label{fig:ntodmap} Real and simulated-noise timestreams for a 
PSB pair over four azimuth half-scans, showing that the noise qualities 
are accurately reproduced.  
The simulated timestreams for all the PSB pairs over the 2 years are 
co-added to form the noise-only maps (right).  
The 1/$f$ noise causes striping in the $T$ map, while the $Q$ polarization 
map approximates white noise. }
\end{figure*} 

(1) For each of the 12 frequency bands of the noise model spectra, we take 
the complex covariance matrix ${\bf {\tilde N}}(f)$ for sums and 
differences of all the good PSB pairs [74$\times$74] and compute its 
complex Cholesky decomposition factor ${\bf L}(f)$ [lower triangular 
74$\times$74] such that ${\bf {\tilde N}}(f) = {\bf L}(f){\bf L}^\dag(f)$.

(2) For each 20-s (200-sample) half-scan and each pair sum or 
difference, we generate the positive-frequency part of a complex spectrum
template ${\boldsymbol \rho}$ [74$\times$100] using normally-distributed 
random numbers whose magnitude has an expectation value of unity.

(3) For each of the 100 positive frequency bins of this vector of 
unit-spectrum templates, we multiply the Cholesky factor from the 
appropriate frequency band of the noise model: ${\bf{\tilde v}}(f)$ = 
${\bf L}(f) {\boldsymbol \rho}(f)$.
The resulting spectra ${\bf{\tilde v}}(f)$ have the same covariance as the 
data:
\begin{eqnarray}
\langle{\bf {\tilde v}}(f){\bf {\tilde v}}^\dag(f)\rangle
&=& \langle{\bf L}(f){\boldsymbol \rho}(f){\boldsymbol \rho}^\dag(f){\bf L}^\dag(f)\rangle \nonumber \\
&=& {\bf L}(f){\bf L}^\dag(f) \nonumber \\
&=& {\bf {\tilde N}}(f).
\end{eqnarray}

(4) To ensure that these ${\tilde{\bf v}}(f)$ approximate the spectra of 
real timestreams, the negative frequency part is set to equal the complex 
conjugate of the positive frequency part, so that the real part is even 
and the imaginary part is odd.

(5) We take the inverse Fourier transform of ${\bf{\tilde v}}(f)$ to 
generate 200 samples of simulated noise time series for each of the 
74 timestreams (sum and difference for the 37 PSB pairs).

To evaluate the accuracy of the noise model and simulation, the 
simulated timestreams were fed back into the noise modeling pipeline and 
the resulting power spectral distributions and covariance matrix were 
compared to those of the real data.  The complex covariance was reproduced 
and the spectral amplitudes agreed to within <1\% rms with no significant 
systematic differences.

The above procedure is used to generate 500 realizations of simulated 
noise timestreams for the entire data set.
As with the real data,
scan-synchronous templates are calculated and subtracted over each set of 
azimuth scans, and the noise timestreams are then co-added into maps.  
Example noise timestreams and maps are illustrated in Figure~\ref{fig:ntodmap}.

\subsection{Noise bias in power spectra}

\begin{figure*}[!htb]
\resizebox{\textwidth}{!}{\includegraphics{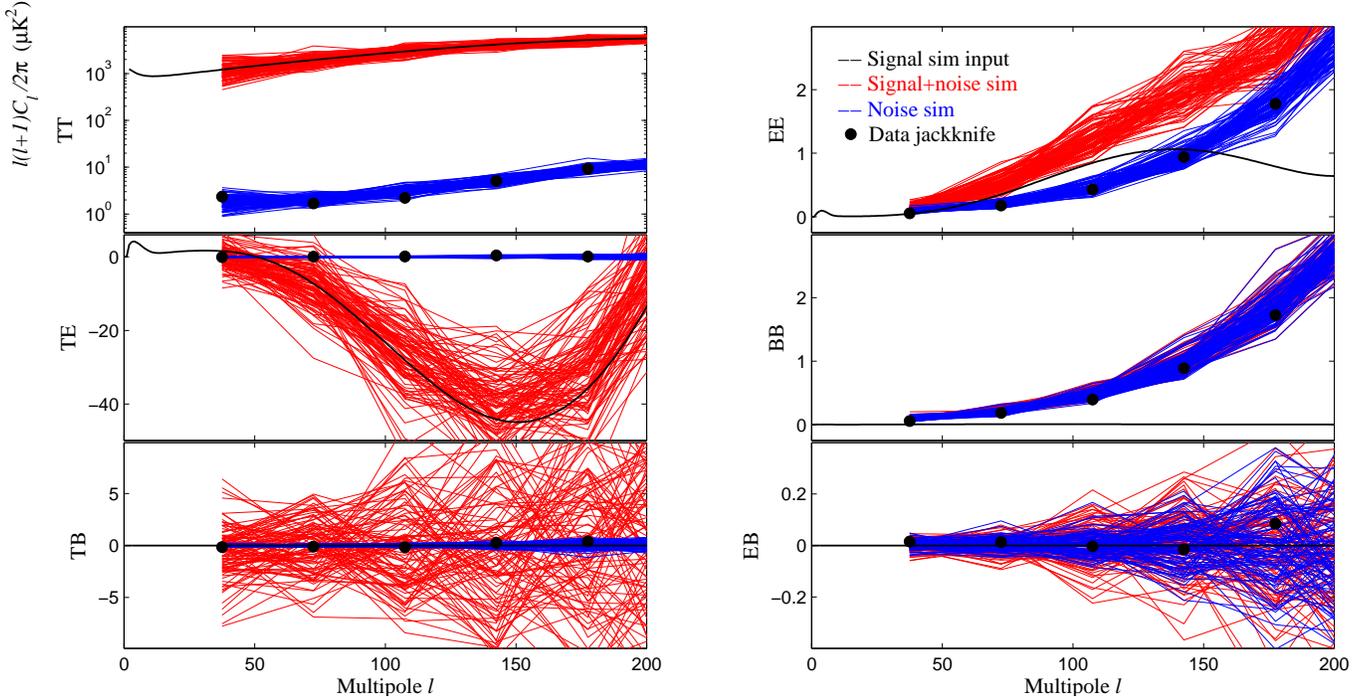}} 
\caption{\label{fig:sn_sim} 
150 GHz power spectra of 100 realizations of simulated signal+noise and noise 
only, compared to the scan-direction jackknife spectra from the actual data.
The distributions of simulated noise spectra are consistent with the data 
jackknife spectra, which are expected to be signal-free and a good 
representation of noise in the data.
The CMB signal simulation uses the input spectra shown.
Error bars in the final spectra are determined by the scatter in the 
signal+noise spectra, which is noise-dominated for $BB$ and $EB$ and 
largely cosmic variance limited for other spectra.  }
\end{figure*}

The noise bias $\langle{\hat N}_\ell\rangle$ is estimated by averaging the 
power spectra from an ensemble of many simulated noise-only maps 
(Figure~\ref{fig:sn_sim}).
The noise bias in $TT$ is 3 orders of magnitude smaller than 
the signal, and the spectrum is sample-variance limited.
For the $TE$, $TB$, and $EB$ cross spectra, the noise from each map is 
mostly uncorrelated, so the resulting $\langle{\hat N}_\ell\rangle$ are 
distributed around zero.
However, the noise contributes a significant portion of the raw $EE$ signal 
and is expected to dominate the $BB$ power spectrum.
The error bar in the final spectrum is based on the scatter in the 
results from an ensemble of signal+noise simulations, and for $BB$ is 
dominated by noise. 

The accuracy of the noise model can be tested by comparing the spectra of the 
simulated noise with those of ``jackknife'' maps, in which two maps made 
with complete halves of the data are differenced so that they are free of CMB 
signal and therefore nominally represent the noise level in the data.
Jackknife divisions included those based on the left/right scan direction 
(shown in Figure~\ref{fig:sn_sim}), azimuth range, boresight rotation 
angle pairs, alternating observing weeks, 2006/2007 observing years, 
and focal plane detector split \citep[more details in][]{Chiang2010}.
We tested for evidence of noise bias amplitude misestimation using these 6 
types of jackknife spectra for $EE$ and $BB$ at 100 and 150 GHz (a total 
of 24 spectra) by comparing the sum of bandpower deviations over 
$\ell=$~300--500 to those from 100 realizations of noise simulations.
The set of jackknife spectra from the actual data are found to be consistent with the simulated 
distributions, even in this high $\ell$ range where the effect of noise 
misestimation is expected to be the largest due to the bias being a 
rapidly increasing function of $\ell$.
Repeating this test with an intentionally introduced $\pm 3\%$
scaling of the noise in the simulations, we find a clearly detectable
departure of the sum of the actual data jackknife bandpower deviations
from those of the simulated distributions.
This allows us to place an upper limit on possible misestimation of
the noise bias at this level, 
at least for a uniformly scaled error across the $\ell$ range.

As described above, a $\pm 3\%$ misestimation of noise power would 
correspond to a maximum shift in our $r$ estimate of 0.1 for the noise 
levels of the current 2-year data set.
The actual 2-year constraint on $r$, as reported in the \citet{Chiang2010}
companion paper, is $r = 0.03^{+0.31}_{-0.27}$.
The jackknife-derived upper limit on possible misestimation of the noise 
bias scales with the noise level. 
Therefore, as the noise in future data releases decreases, we can expect this
internal jackknife test to continue to allow us to place upper limits on noise misestimation 
(or to detect it if present) at a level corresponding to roughly 1/3 of the total 
statistical uncertainty on $r$, assuming a noise-limited $BB$ spectrum,
and less than this if a $BB$ signal is detected.
The final results of power spectrum estimation with noise bias removal and error bars 
based on simulations of signal and noise for the \bicep\ 2-year data are
presented in the CMB results paper.

\section{Conclusion}

\bicep\ is an experiment built with a primary goal of targeting 
the signature of inflationary gravitational waves
in the $B$-mode polarization of the CMB at 
angular scales near the expected peak around 2$\deg$.
Its novel design emphasizes simplicity and systematic control,
employing a carefully baffled compact cryogenic refractor
and relying on a simple observing strategy of azimuth-scan modulation
with periodic boresight rotation.
Using \bicep's actual data analysis pipelines, we have identified 
those aspects of the experiment's instrumental and noise properties
which require careful control and characterization.
We have established benchmarks for these quantities corresponding to 
the expected $B$-mode polarization signal for a tensor-to-scalar ratio 
of $r=0.1$, a value several times smaller than the level of statistical
uncertainty of the \bicep\ 2-year result, 
$r = 0.03^{+0.31}_{-0.27}$, or $r < 0.72$ at 95\% confidence
\citep{Chiang2010}.
  
The instrumental characterization reported in this paper shows that all
studied sources
of potential systematic errors except for the pair relative gains,
differential pointing, and possibly the noise estimation
contribute to 
the uncertainty in the measurement at a level of $r\lesssim0.01$.
The effects which were found to be controlled to this 
level include differential beam size, differential ellipticity, polarization 
orientation uncertainty, telescope pointing, sidelobes, and thermal 
stability.  In addition, effects which impact the overall calibration
of the polarization spectra, including the absolute gain, cross-polarization
response, and relative polarization orientation, have been characterized
well enough to ensure that calibration uncertainty is a small fraction of
our error budget.

Of the remaining three effects, the noise is estimated with sufficient accuracy to limit the
uncertainty contribution to $r<0.1$, a number that will improve as
more data are added. 
The differential pointing of the PSB pairs is sufficiently small to meet 
our $r=0.1$ benchmark without correction, but will need to be 
taken into account to achieve lower limits on $r$. 
The current uncertainty in the method we use to verify our calibration of
relative detector gains in the PSB pairs leads to an upper limit in
possible error on $r$ which slightly exceeds our $r=0.1$ benchmark.
The uncertainty will improve as more data are included in the analysis, and 
a different approach may be needed if 
any significant relative gain errors are measured.   
By employing a more sophisticated analysis which
allows for imperfectly-corrected differential gains of the 
PSB pairs and which includes the measured
differential pointing,
we expect to be able to control all studied sources of potential
systematic errors to levels far below $r=0.1$.

This practical experience with \bicep\ provides a guide for future 
experiments in search for the signature of inflationary 
gravitational waves in CMB polarization. 

\acknowledgements
\bicep\ is supported by NSF Grant OPP-0230438, Caltech Discovery Fund, 
Caltech President's Fund PF-471, JPL Research and Technology Fund, and the 
late J.~Robinson.  
We thank our colleagues in \acbar, \boom, \QUAD, \bolocam, and \spt\ for 
advice and helpful discussions, Kathy Deniston for logistical and 
administrative support, and the South Pole Station staff for their 
support. We acknowledge support by the NASA Graduate Fellowship program 
(H.C.C., E.M.B.), the John~B. and Nelly Kilroy Foundation (J.M.K.), and 
NSF PECASE Award AST-0548262 (B.G.K.). \\

\bibliographystyle{apj}
\bibliography{2009_calibration_noise}

\end{document}